\documentclass[aps,prb,superscriptaddress,twocolumn,10pt,floatfix,dvipsnames]{revtex4-2}
\usepackage[utf8]{inputenc}
\usepackage{amsfonts}
\usepackage{amssymb}
\usepackage{amsmath}
\usepackage{mathtools,units,bm}
\usepackage{mathptmx}
\usepackage{microtype}
\usepackage{graphicx}
\usepackage{cancel}
\usepackage{orcidlink}
\usepackage{tikz}

\usetikzlibrary{arrows.meta,decorations.markings}
\usepackage[normalem]{ulem}
\usepackage{soul} % para \st (strikeout)
\usepackage{hyperref}
\hypersetup{colorlinks=true,citecolor=blue,linkcolor=blue,filecolor=blue,urlcolor=blue,pdfstartview=FitH,pdfpagemode=UseNone,pdfauthor={},pdfdisplaydoctitle=true,pdfduplex=DuplexFlipLongEdge}

\emergencystretch=\maxdimen
\usepackage{lipsum}

\begin{document}
\normalem

\title{Mott transitions in Dirac nodal-line semimetals}

\author{Andressa R.~Medeiros-Silva~\orcidlink{0000-0001-7532-5529}}
\affiliation{International Institute of Physics, Universidade Federal do Rio Grande do Norte, Natal, RN 59078-970, Brazil}
\author{Lucas O. Lima~\orcidlink{0000-0001-8319-1495}}
\affiliation{Instituto de F\'isica, Universidade Federal de Goi\'as, Goi\^ania, GO 74690-900, Brazil}
\author{Marcello~B.~da~Silva~Neto~\orcidlink{0000-0002-6817-3472}}
\affiliation{Instituto de F\'isica, Universidade Federal do Rio de Janeiro, Rio de Janeiro, RJ 21941-972, Brazil}
\author{Fakher F. Assaad~\orcidlink{0000-0002-3302-9243}}
\affiliation{Institut f\"ur Theoretische Physik und Astrophysik, Universit\"at W\"urzburg, 97074 W\"urzburg, Germany}
\affiliation{W\"urzburg-Dresden Cluster of Excellence ctd.qmat, Am Hubland, 97074 W\"urzburg, Germany}
\author{Natanael C. Costa~\orcidlink{0000-0003-4285-4672}}
\affiliation{Institut f\"ur Theoretische Physik und Astrophysik, Universit\"at W\"urzburg, 97074 W\"urzburg, Germany}
\affiliation{Instituto de F\'isica, Universidade Federal do Rio de Janeiro, Rio de Janeiro, RJ 21941-972, Brazil}

\begin{abstract}
Nodal-line semimetals provide a natural setting to investigate how the geometry of a nodal manifold affects interaction-driven quantum phase transitions. Here, we study the Hubbard model on a cubic lattice supporting either nodal-point (cubic $\pi$-flux model) or nodal-line semimetallic states using mean-field theory, determinant quantum Monte Carlo simulations, and a weak-coupling renormalization-group (RG) analysis. In both cases, antiferromagnetic order develops only above a finite interaction strength, with the nodal-line system exhibiting a smaller critical coupling due to its enhanced low-energy density of states and markedly stronger finite-size effects. For the nodal-point model, the quantum critical behavior is compatible with $(3+1)$-dimensional Gross--Neveu mean-field scaling. In contrast, the weak-coupling RG reveals that the nodal line itself is unstable against interactions: although the Hubbard interaction is perturbatively irrelevant, its second-order self-energy generates a symmetry-allowed relevant term proportional to $U^2\cos k_z$, reconstructing the nodal line into isolated, strongly anisotropic nodal points. This weak-coupling reconstruction is distinct from the strong-coupling antiferromagnetic transition and highlights the role of nodal-manifold geometry in interacting semimetals.
\end{abstract}

\maketitle

%%%%%%%%%%%%%%%%%%%%%%%%%%%%%%%%%%%%%%%%%%%%%%%%%%
\section{Introduction}
%%%%%%%%%%%%%%%%%%%%%%%%%%%%%%%%%%%%%%%%%%%%%%%%%%

Topological semimetals have attracted considerable attention as systems in which the low-energy electronic structure is governed by symmetry-protected band touchings. In Dirac and Weyl semimetals, the conduction and valence bands touch at isolated points in the Brillouin zone, giving rise to quasiparticles with linear dispersion around the nodal points\,\cite{CastroNeto2009,Wehling2014}. In three dimensions, Dirac semimetals may appear either at the transition between topologically distinct insulating phases or as stable semimetallic phases protected by crystalline symmetries \cite{young2012,Armitage2018}. The pointlike character of the Fermi surface leads to a density of states that vanishes as $\rho(\omega)\sim \omega^2$, rendering weak short-range interactions perturbatively irrelevant\,\cite{Maciejko2014}.

Despite this weak-coupling stability, three-dimensional Dirac systems may display nontrivial many-body and disorder-driven phenomena. Long-range Coulomb interactions can renormalize quasiparticle properties and, at sufficiently strong coupling, drive dynamical mass generation or excitonic instabilities \cite{gonzalez2015,wang2020}. Short-range interactions may also destabilize the semimetal and generate scalar, pseudoscalar, or axionic insulating masses \cite{roy2016}. In addition, disorder can induce transitions from an incompressible Dirac semimetal to a compressible diffusive metal and, at stronger disorder, to an Anderson insulating phase \cite{pixley2015,roy2016,fu2017,santospires2021}. These results establish three-dimensional Dirac semimetals as a useful toy model for studying the interplay between topology, correlations, and randomness.

A particularly useful lattice realization of three-dimensional Dirac fermions is the cubic $\pi$-flux model\,\cite{hosur2010}. In this system, each plaquette encloses a $\pi$ flux, and the noninteracting spectrum hosts linearly dispersing Dirac cones. The Hubbard model on this lattice has been investigated using determinant quantum Monte Carlo  simulations, which showed that the Dirac semimetal remains stable against weak onsite repulsion and undergoes a direct transition to an antiferromagnetic insulating phase only above a finite critical interaction \cite{huang2020}. 
For three-dimensional Dirac fermions, the corresponding
Gross--Neveu--Yukawa theory lies at its upper critical
spacetime dimension $d_c=4$, such that the critical behavior
is Gaussian up to logarithmic corrections \cite{Moshe2003}.

Nodal-line semimetals provide a natural extension of this setting. In these systems, the conduction and valence bands touch not at isolated points, but along one-dimensional manifolds in momentum space, which may form loops, chains, or more complex structures \cite{fang2016,liu2021}. The stability of these nodal lines is generally tied to crystalline or internal symmetries, such as time-reversal, inversion, mirror, or glide symmetry \cite{Malard22_SciRep}. When the protecting symmetry is broken, the nodal line may become fully gapped or evolve into other topological band structures. Compared with a three-dimensional Dirac cone, a nodal-line semimetal has a larger low-energy phase space. Therefore, interaction effects are expected to be enhanced relative to point-node Dirac systems, although the vanishing density of states at the Fermi level still suggests that a stable weak-coupling semimetallic regime may survive.
Indeed, related studies of excitonic quantum criticality in ZrSiS suggest that nodal-line systems may also display strong quasiparticle mass renormalization near a semimetal--excitonic-insulator quantum critical point, even in regimes where a full gap is not observed experimentally \cite{wang2020}.

The role of electronic interactions in nodal-line semimetals is therefore a natural and timely question \cite{Huh2016,Roy2017,Kang2019,Medeiros-Silva25}. Field-theoretical studies based on renormalization-group and Gross--Neveu--Yukawa approaches have shown that short-range interactions can drive continuous quantum phase transitions from a nodal-line semimetal into fully gapped phases in charge, spin, and superconducting channels \cite{jose2020}. In these approaches, the many-body instability occurs only beyond a finite interaction strength, and the critical exponents depend on the number of order-parameter components. In particular, the dynamical critical exponent is found to be $z=1$ at one-loop order, while the remaining exponents acquire corrections associated with the anomalous dimension of the bosonic order parameter \cite{jose2020}. 

More broadly, studies of anisotropic and semi-Dirac semimetals emphasize that the geometry of the band touching plays a central role in determining the effects of interactions. Systems with mixed linear and quadratic dispersions interpolate between relativistic Dirac fermions and quadratic band-touching systems, leading to modified density of states scaling and distinct quantum critical behavior \cite{sur2019,uryszek2020,wang2023semidirac}. In some cases, interactions can generate additional terms in the low-energy dispersion, modify the stability of the noninteracting fixed point, or promote several competing ordered phases \cite{Roy2017,Kang2019,pujari2016,wang2023semidirac,elsayed2025}. These results indicate that the universality class of an interaction-driven transition cannot be inferred from spatial dimensionality alone, but depends sensitively on the structure of the nodal manifold and on the dispersion normal to it.

\begin{figure}[t]
\centering
\includegraphics[width=\linewidth]{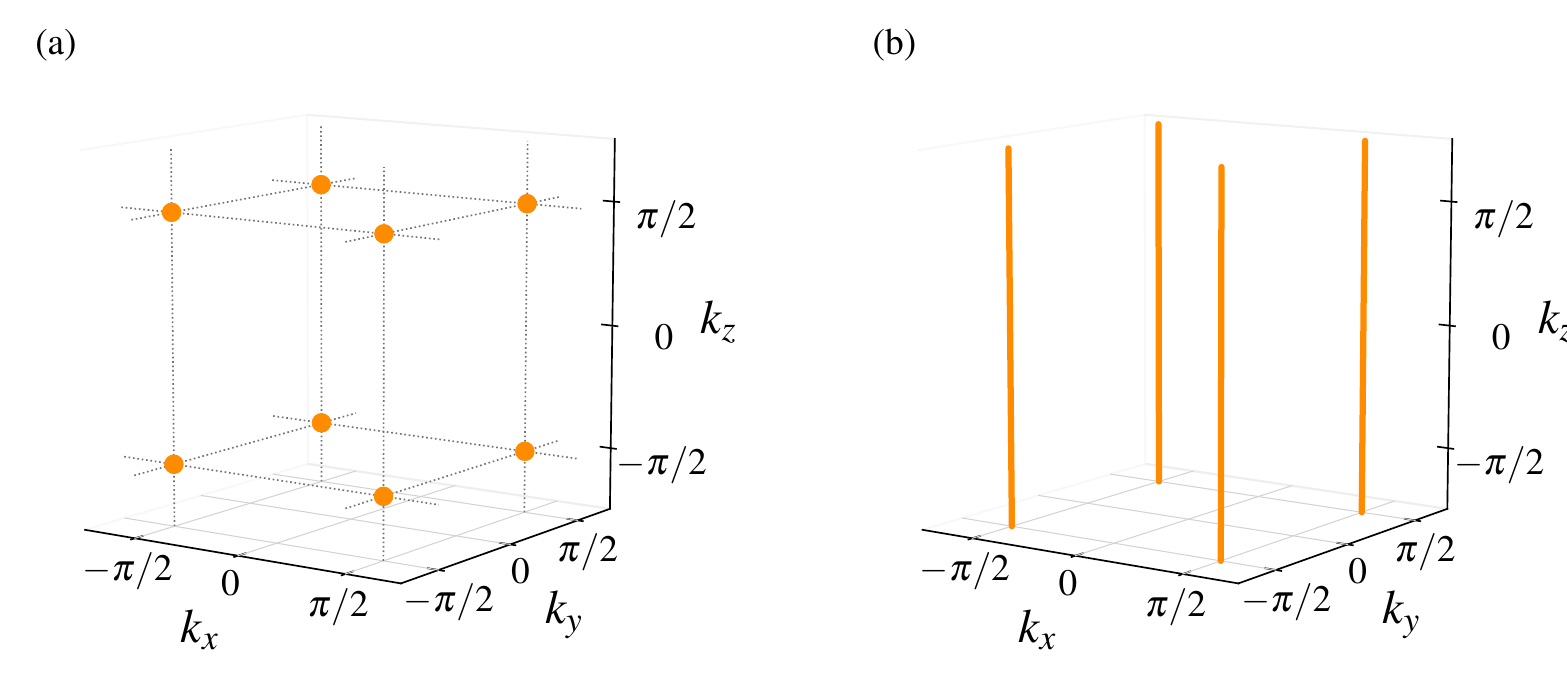} 
\caption{Noninteracting Fermi surface at half filling for (a) the cubic $\pi$-flux model and (b) the corresponding nodal-line model, with nodal lines located at $\left(\pm\frac{\pi}{2},\pm\frac{\pi}{2},k_z\right)$.} 
\label{fig:FS}
\end{figure}

Motivated by these developments, we construct a nodal-line semimetal model starting from the cubic $\pi$-flux lattice. By adding hopping terms to the parent Dirac model, we generate a continuous nodal line along $(\pi/2,\pi/2,k_z)$ in the Brillouin zone, as depicted in Fig.\,\ref{fig:FS}. This construction allows us to compare two closely related systems: the original cubic $\pi$-flux Dirac semimetal, and a modified nodal-line semimetal. The central questions are whether the nodal line remains stable against weak interactions, whether antiferromagnetic order appears only above a finite critical coupling, and whether the corresponding critical behavior differs from that of the three-dimensional Dirac case.

In this work, we combine mean-field theory, quantum Monte Carlo simulations, and a weak-coupling renormalization-group analysis to study the Hubbard model on both the cubic $\pi$-flux lattice and the modified nodal-line system. We first benchmark our calculations in the parent Dirac semimetal by examining the antiferromagnetic critical interaction, magnetic correlations, and finite-size scaling behavior. We then apply the same analysis to the nodal-line model in order to determine the stability of the semimetallic phase. In this way, our study clarifies how the Hubbard-driven quantum phase transition is modified when isolated Dirac points are promoted to an extended nodal line. Finally, a weak-coupling RG analysis reveals a distinction between the stability of the semimetal against symmetry-breaking order and the stability of the nodal manifold itself. 
The paper is organized as follows. In the next section, we introduce the model and describe the methodologies employed. The results for both the nodal-point and nodal-line cases are presented in Sec.~\ref{Sec:Results}, while our conclusions are summarized in Sec.~\ref{sec:conclusions}. 

%%%%%%%%%%%%%%%%%%%%%%%%%%%%%%%%%%%%%%%%%%%%%%%%%%
\section{Model and methodology}
%%%%%%%%%%%%%%%%%%%%%%%%%%%%%%%%%%%%%%%%%%%%%%%%%%
\label{Sec: Model}

\subsection{The model}

We investigate interacting fermions in a three-dimensional lattice model for nodal-point and nodal-line semimetals. The noninteracting Hamiltonian (in momentum space) reads
\begin{align}
\mathcal{H}_{0}
=&
\ 2t
\sum_{\mathbf{k},\sigma}
\sum_{\tau,\mu}
\sum_{\tau',\mu'}
c^{\dagger}_{\mathbf{k},\tau,\mu,\sigma}
\biggl[
(\tau_x)_{\tau\tau'}\delta_{\mu\mu'}\cos k_x
\nonumber\\
&+
(\tau_y)_{\tau\tau'}(\mu_y)_{\mu\mu'}\cos k_y
+
(\tau_z)_{\tau\tau'}\delta_{\mu\mu'}\chi(\mathbf{k})
\biggr]
c_{\mathbf{k},\tau',\mu',\sigma}\ .
\label{eq:H_toy}
\end{align}
Here \(\tau_i\) and \(\mu_i\) are Pauli matrices acting on two independent orbital degrees of freedom, and \(\sigma\) denotes the spin.
This general Hamiltonian is useful because it allows us to interpolate between nodal-point (NP) and nodal-line (NL) regimes by choosing, respectively,
\begin{equation}
\chi(\mathbf{k}) =
\begin{cases}
\chi_{\rm NP}(\mathbf{k}) = -\cos k_z ~,\\[4pt]
\chi_{\rm NL}(\mathbf{k}) = -\cos k_z \left( \cos^2 k_x+\cos^2 k_y \right) ~.
\end{cases}
\end{equation}
As discussed below, for $\chi=\chi_{\rm NP}$ the spectrum exhibits nodal/Dirac points in the Brillouin zone, as shown in Fig.\,\ref{fig:FS}\,(a). By contrast, for $\chi=\chi_{\rm NL}$, the band touching extends into a nodal line along the $k_z$ direction, as shown in Fig.\,\ref{fig:FS}\,(b).

For convenience, we combine the two orbital labels $(\tau,\mu)$ into a single index $\alpha=A,B,C,D$, according to
$(0,0)\equiv A$,
$(0,1)\equiv B$,
$(1,0)\equiv C$,
$(1,1)\equiv D$.
Within this notation, Eq.\,\eqref{eq:H_toy} becomes
$
\mathcal{H}_0 =
\sum_{\mathbf{k},\sigma}
\Psi^\dagger_{\mathbf{k}\sigma}
\hat{H}_0(\mathbf{k})
\Psi_{\mathbf{k}\sigma},
$
where
\begin{equation}
\hat{H}_{0}(\mathbf{k}) = 2t
\begin{pmatrix}
    \chi(\mathbf{k}) & 0 & \cos k_x & -\cos k_y\\
    0 & \chi(\mathbf{k}) & \cos k_y & \cos k_x\\
    \cos k_x & \cos k_y & -\chi(\mathbf{k}) & 0\\
    -\cos k_y & \cos k_x & 0 & -\chi(\mathbf{k})
\end{pmatrix}~,
\label{eq:H_toy_matrix}
\end{equation}
with $\Psi_{\mathbf{k}\sigma}=
(c_{A,\mathbf{k}\sigma},
c_{B,\mathbf{k}\sigma},
c_{C,\mathbf{k}\sigma},
c_{D,\mathbf{k}\sigma})^T$ being a four-component spinor in orbital space. 
Regardless of the choice of $\chi(\mathbf{k})$, the corresponding energy bands are doubly degenerate and given by
\begin{equation}
E_{\pm,\mathbf{k}}
=
\pm 2t
\sqrt{
\cos^2 k_x
+
\cos^2 k_y
+
\chi^2(\mathbf{k})
}.
\label{eq:bands_toy}
\end{equation}
Therefore, band touchings occur when
$\cos k_x=\cos k_y=\chi(\mathbf{k})=0$.
For $\chi_{\rm NL}$, this condition is satisfied along the line
$(k_x,k_y)=(\pi/2,\pi/2)$, with arbitrary $k_z$, while for $\chi_{\rm NP}$, the last condition reduces to $\cos k_z=0$, producing Dirac points at $(k_x,k_y,k_z)=(\pi/2,\pi/2,\pi/2)$, as shown in Fig.\,\ref{fig:FS}.

\begin{figure}[t]
		\centering
        \includegraphics[width=0.7\linewidth]{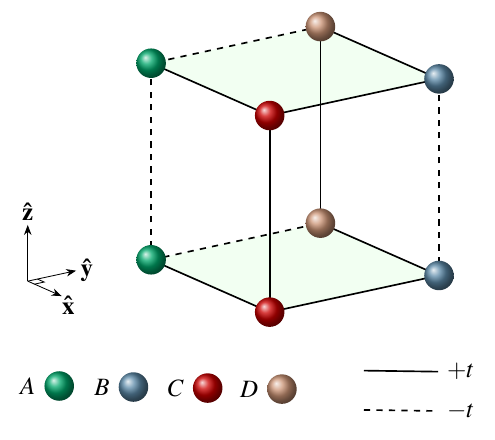} 
        \caption{Real-space representation of the cubic $\pi$-flux model. The four orbitals within the unit cell are labeled $A$, $B$, $C$, and $D$. Solid and dashed bonds denote nearest-neighbor hopping amplitudes $+t$ and $-t$, respectively, arranged such that each elementary plaquette is threaded by a $\pi$ flux.}
        \label{fig:lattice}
\end{figure}

In this representation, the noninteracting Hamiltonian reads
\begin{equation}
\hat H_0(\mathbf k)=2t\left[
\cos k_x\,\Gamma_1+
\cos k_y\,\Gamma_2+
\chi(\mathbf k)\,\Gamma_3
\right],
\label{eq:H_toy2}
\end{equation}
where
\begin{equation}
\begin{aligned}
\Gamma_1&=\tau_x\otimes\mu_0,&
\Gamma_2&=\tau_y\otimes\mu_y,\\
\Gamma_3&=\tau_z\otimes\mu_0,&
\Gamma_4&=\tau_y\otimes\mu_x,\\
\Gamma_5&=\tau_y\otimes\mu_z.
\end{aligned}
\label{eq:Gamma_toy_QMC}
\end{equation}
These matrices satisfy the Clifford algebra $\{\Gamma_a,\Gamma_b\}=2\delta_{ab}I_4$, with $a,b=1,\ldots,5$.

The real-space representation of the Hamiltonian in Eqs.\,\eqref{eq:H_toy} or \eqref{eq:H_toy2} is given in Appendix~\ref{app:real_space}, using the primitive vectors
\begin{equation}
\mathbf a_1=(2,0,0)a,\quad
\mathbf a_2=(0,2,0)a,\quad
\mathbf a_3=(0,0,1)a,
\label{Eq:privectors}
\end{equation}
with $a$ being the lattice spacing.
The NP limit corresponds to the three-dimensional cubic $\pi$-flux Dirac model\,\cite{hosur2010}, as displayed in Fig.\,\ref{fig:lattice}, while the NL limit is obtained by adding extra hopping processes to this parent NP Hamiltonian.
Some noninteracting properties of the NL case are discussed in Appendix~\ref{app:dos}.
Hereafter, we measure energies in units of the hopping amplitude $t$ and set the lattice constant to unity.

%%%%%%%%%%%%%%%%%%%%%%%%%%%%%%%%%%%%%%%%%%%

\subsection{The mass term}
\label{sucsec:mass}

Despite its simplicity, Eq.\,\eqref{eq:H_toy2} does not make explicit the mass term that appears when antiferromagnetic (AFM) order emerges.
From the real-space representation (see Appendix~\ref{app:real_space}),
one can alternatively define the primitive vectors as
\begin{equation}
\widetilde{\mathbf a}_1=(1,1,0)a,\quad
\widetilde{\mathbf a}_2=(1,-1,0)a,\quad
\widetilde{\mathbf a}_3=(0,0,2)a~.
\end{equation}
For this choice, the orbital positions inside the unit cell are
\begin{equation}
\begin{aligned}
\boldsymbol\delta_A&=(0,0,0),&
\boldsymbol\delta_C&=(1,0,0),\\
\boldsymbol\delta_{A'}&=(0,0,1),&
\boldsymbol\delta_{C'}&=(1,0,1),
\end{aligned}
\end{equation}
and a local gauge transformation $c_{\mathbf r\sigma}=q_y d_{\mathbf r\sigma}$, with $q_y=1$ for even $y$ and $q_y=i$ for odd $y$, is required to make the hopping amplitudes periodic.

Performing a Fourier transform, the noninteracting Hamiltonian reads
\begin{equation}
\mathcal H_0=\sum_{\mathbf k,\sigma}
\widetilde\Psi^\dagger_{\mathbf k\sigma}
\widetilde H_0(\mathbf k)
\widetilde\Psi_{\mathbf k\sigma},
\end{equation}
with
\begin{equation}
\widetilde H_0(\mathbf k)=2t\left[
\cos k_x\,\widetilde\Gamma_1+
\cos k_y\,\widetilde\Gamma_2+
\chi(\mathbf k)\,\widetilde\Gamma_3
\right],
\label{eq:H_toy_clifford}
\end{equation}
where $\widetilde\Psi_{\mathbf k\sigma}=(d_{A,\mathbf k\sigma},d_{C,\mathbf k\sigma},d_{A',\mathbf k\sigma},d_{C',\mathbf k\sigma})^T$ and $\sigma$ labels the electron spin. Similarly, the matrices
\begin{equation}
\begin{aligned}
\widetilde\Gamma_1&=\widetilde\mu_0\otimes\widetilde\tau_x,&
\widetilde\Gamma_2&=\widetilde\mu_0\otimes\widetilde\tau_y,\\
\widetilde\Gamma_3&=\widetilde\mu_x\otimes\widetilde\tau_z,&
\widetilde\Gamma_4&=\widetilde\mu_z\otimes\widetilde\tau_z,\\
\widetilde\Gamma_5&=\widetilde\mu_y\otimes\widetilde\tau_z
\end{aligned}
\label{eq:Gamma_toy}
\end{equation}
satisfy the Clifford algebra
$\{\widetilde\Gamma_a,\widetilde\Gamma_b\}=2\delta_{ab}I_4$.
Here $\widetilde\mu_i$ and $\widetilde\tau_i$ are Pauli matrices acting on two independent orbital indices: $\widetilde\mu_i$ distinguishes the unprimed layer $(A,C)$ from the primed layer $(A',C')$, while $\widetilde\tau_i$ distinguishes the $A$ and $C$ sites within each layer, with $\widetilde\mu_0=\widetilde\tau_0=I_2$.

Within this representation, a staggered antiferromagnetic pattern is given by a term $\sigma \Delta \widetilde\Gamma_4$, where $\sigma=\pm1$, and $\Delta$ is proportional to the magnetic order parameter, with ordering wavevector $\widetilde{\mathbf Q}=0$. Since
$\{\widetilde\Gamma_4,\widetilde H_0(\mathbf k)\}=0$,
this contribution is a fermion mass, making explicit the coupling between the AFM order parameter and the Dirac fermions, as in Gross--Neveu systems.

Nevertheless, hereafter, we adopt the unit cell defined in Eq.\,\eqref{Eq:privectors}, retaining the original hopping gauge. Although this choice makes the AFM mass term less explicit, it is particularly convenient for quantum Monte Carlo simulations. It changes the orbital grouping and the corresponding Brillouin zone without altering the underlying lattice model.

\subsection{Mean-field approximation}
\label{sucsec:MFT}

%%%%%%%%%%%%%%%%%%%%%%%%%%%%%%%%%%%%%%%%%%%

We now include the onsite Hubbard interaction,
\begin{equation}
\mathcal{H}_U =
U\sum_{\mathbf{R},\alpha}
n_{\mathbf{R}_{\alpha}\uparrow}
n_{\mathbf{R}_{\alpha}\downarrow}.
\end{equation}
As outlined in Appendix \ref{app:MFT}, within a static mean-field approximation, assuming an AFM order, one obtains
the dispersion
$E_{\pm,\mathbf{k},\sigma}(\Delta)
=
\pm \varepsilon_{\mathbf{k},\sigma}(\Delta)$, 
where
\begin{equation}
\varepsilon_{\mathbf{k},\sigma}(\Delta)
=
2t
\sqrt{
\cos^2 k_x+
\cos^2 k_y+
\chi^2(\mathbf{k})
+
\Delta^2
}~,
\label{Eq:espectrum}
\end{equation}
with $\Delta=\frac{U m}{4t}$, and $m$ being the AFM order parameter.
Here, the staggered AFM order is characterized by the ordering wave vector $\mathbf Q=(0,0,\pi)$ in the extended Brillouin zone. 

Given this, the mean-field free energy is obtained by summing over the quasiparticle levels,
\begin{equation}
F(\Delta)
=
-\frac{8}{\beta}
\sum_{\mathbf{k},\sigma}
\sum_{p=\pm}
\ln\left[
1+
e^{-\beta E_{p,\mathbf{k},\sigma}(\Delta)}
\right]
+
64 N\frac{t^2}{U}\Delta^2,
\label{eq:MF_free_energy}
\end{equation}
with $\Delta$ being obtained self-consistently by minimizing $F(\Delta)$.
That is, by employing $\partial F(\Delta) / \partial \Delta = 0$ at zero temperature, one obtains
\begin{equation}
\frac{t}{U}
=
\frac{1}{4N}
\sum_{\mathbf{k}}
\frac{1}{
\sqrt{\cos^2 k_x + \cos^2 k_y + \chi^2(\mathbf{k}) + \Delta^2}
}~,
\label{eq:MF_order_parameter}
\end{equation}
which provides the order parameter $m$.
The critical interaction strength $U_c$ is obtained by taking the limit
$\Delta\rightarrow 0$,
\begin{equation}
\frac{t}{U_c} = \frac{1}{4N} \sum_{\mathbf{k}}
\frac{1}{\sqrt{ \cos^2 k_x + \cos^2 k_y + \chi^2(\mathbf{k})}}~.
\label{eq:gap_T0}
\end{equation}

\subsection{Determinant quantum Monte Carlo method}

We now turn to the determinant quantum Monte Carlo (DQMC) simulations \cite{blankenbecler81,hirsch85,white89,dosSantos03b}. This method provides a numerically exact treatment of the interacting lattice problem, up to statistical errors and the controllable Trotter--Suzuki discretization error. 

Considering our Hamiltonian as the sum of a quadratic kinetic term, $\hat K\equiv \mathcal{H}_{0}$, and the interaction term, $\hat P \equiv \mathcal{H}_{U}$, we perform a Trotter--Suzuki decomposition in the partition function,
\begin{align}
Z
&=
{\rm Tr}\, e^{-\beta \mathcal{H}}
\nonumber\\
&=
{\rm Tr}
\left[
e^{-\Delta\tau(\hat K+\hat P)}
\right]^{L_{\tau}}
\nonumber\\
&\approx
{\rm Tr}
\left[
e^{-\Delta\tau \hat K}
e^{-\Delta\tau \hat P}
\right]^{L_{\tau}} .
\label{eq:TS_decoupling}
\end{align}
That is, we write the partition function as an imaginary-time path integral by discretizing the interval $\tau \in[0,\beta)$ into $L_{\tau}$ slices of width $\Delta\tau$, with $\beta=L_{\tau}\Delta\tau$ being the inverse temperature. 
The decomposition in Eq.\,\eqref{eq:TS_decoupling} introduces systematic errors of order $(\Delta\tau)^2$ in measured observables. In our simulations we use $\Delta\tau\leq 0.1$, for which the remaining discretization effects are smaller than, or comparable to, the statistical uncertainties.

The quartic interaction term is then decoupled by a discrete Hubbard--Stratonovich transformation \cite{Hirsch83}. For each space-time point $(\mathbf{R}_{\alpha},\tau)$, we introduce an auxiliary Ising field
$s_{\mathbf{R}_{\alpha},\tau}=\pm1$ through
\begin{equation}
e^{-\Delta\tau U
\left(n_{\uparrow}-\frac{1}{2}\right)
\left(n_{\downarrow}-\frac{1}{2}\right)}
=
\frac{1}{2}
e^{-\Delta\tau U/4}
\sum_{s=\pm1}
e^{\lambda s (n_{\uparrow}-n_{\downarrow})},
\label{eq:HS_transformation}
\end{equation}
where the coupling $\lambda$ is defined by
$\cosh\lambda=e^{\Delta\tau U/2}$.
After this transformation, the fermionic degrees of freedom appear only quadratically and can be traced out exactly for a fixed auxiliary-field configuration. The partition function becomes
\begin{equation}
Z=
\sum_{\{s\}}
W(\{s\}),
\qquad
W(\{s\})
=
\det M_{\uparrow}(\{s\})
\det M_{\downarrow}(\{s\}),
\end{equation}
where $M_{\sigma}$ is the fermionic matrix associated with spin sector $\sigma$ in the background of the Hubbard--Stratonovich field. Physical observables are obtained by sampling the auxiliary fields with probability proportional to $W(\{s\})$.

In general, DQMC simulations of fermionic systems may suffer from the minus-sign problem when the product of determinants is not positive \cite{Loh90,troyer2005computational}. In the present work, however, we restrict ourselves to the half-filled case, where particle-hole symmetry ensures sign-problem-free simulations for the models considered here. This allows us to access the low-temperature regime needed to analyze the onset of antiferromagnetic order and the stability of the semimetallic phase.

For each set of parameters, we perform between 10 and 30 independent DQMC runs initialized with different random seeds. In each run, we discard $0.8$--$2\times10^3$ sweeps for thermalization and subsequently accumulate measurements over $0.6$--$3\times10^3$ sweeps, depending on temperature and lattice size.  These simulation parameters were chosen to ensure proper equilibration and reliable ensemble-averaged results.
We also average over periodic (PBC) and antiperiodic (APBC) boundary conditions, which reduces finite-size effects for thermodynamic quantities and correlation functions.

%%%%%%%%%%%%%%%%%%%%%%%%%%%%%%%%%%%%%%%%%%%%%%%%%%
\section{Results}
%%%%%%%%%%%%%%%%%%%%%%%%%%%%%%%%%%%%%%%%%%%%%%%%%%
\label{Sec:Results}

\subsection{Noninteracting and mean-field cases}

\begin{figure}[t]
\centering 
\includegraphics[width=\linewidth]{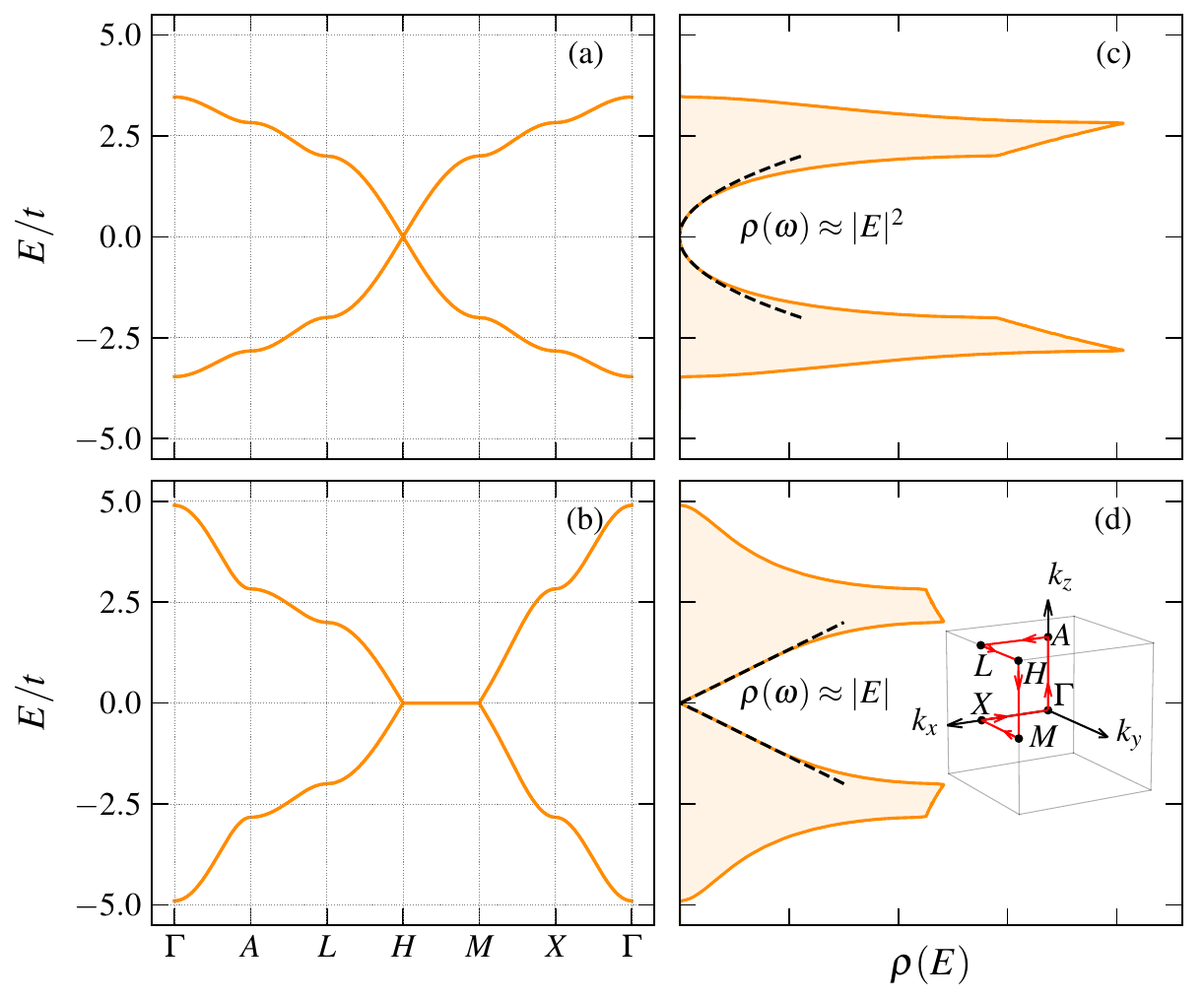} 
\caption{(a)\;Dispersion along the high-symmetry path and (c)\;density of states for nodal points. (b) and (d) show the same for nodal lines.} 
\label{fig:DOS_per_site}
\end{figure}

We first compare the NP and NL limits at the noninteracting level. Figures~\ref{fig:DOS_per_site}\,(a) and \ref{fig:DOS_per_site}\,(b) show the dispersion relation of Eq.\,\eqref{eq:bands_toy} along high-symmetry directions for the NP and NL cases, respectively. In the nodal-point case, the spectrum exhibits a Dirac cone centered at $(\pi/2,\pi/2,\pi/2)$. By contrast, in the NL case, the bands touch along the $k_z$ direction when $k_x=k_y=\pi/2$, producing a dispersionless nodal feature as a function of $k_z$.

This difference in the low-energy band structure leads to distinct densities of states (DOS). As shown in Appendix~\ref{app:dos}, the low-energy DOS can be obtained from the continuum expansion around the corresponding nodal manifold. For the NP case, one finds
\begin{equation}
\rho_{\rm NP}(E)
=
\frac{E^2}{2\pi^2t^3},
\label{eq:DOSNP}
\end{equation}
which shows that the DOS vanishes quadratically at the nodal point.
In contrast, the NL case gives
\begin{equation}
\rho_{\rm NL}(E)
=
\frac{|E|}{2\pi t^2},
\label{eq:DOSNL}
\end{equation}
corresponding to a DOS that vanishes linearly with energy.
This behavior is confirmed by the numerical DOS shown in
Figs.\,\ref{fig:DOS_per_site}\,(c) and \ref{fig:DOS_per_site}\,(d)
for the NP and NL cases, respectively. The DOS
analysis is useful because the different low-energy power laws provide the first indication that the two systems may respond differently once electronic interactions are included, as discussed below.

\begin{figure}[t]
\centering
\includegraphics[width=0.99\linewidth]{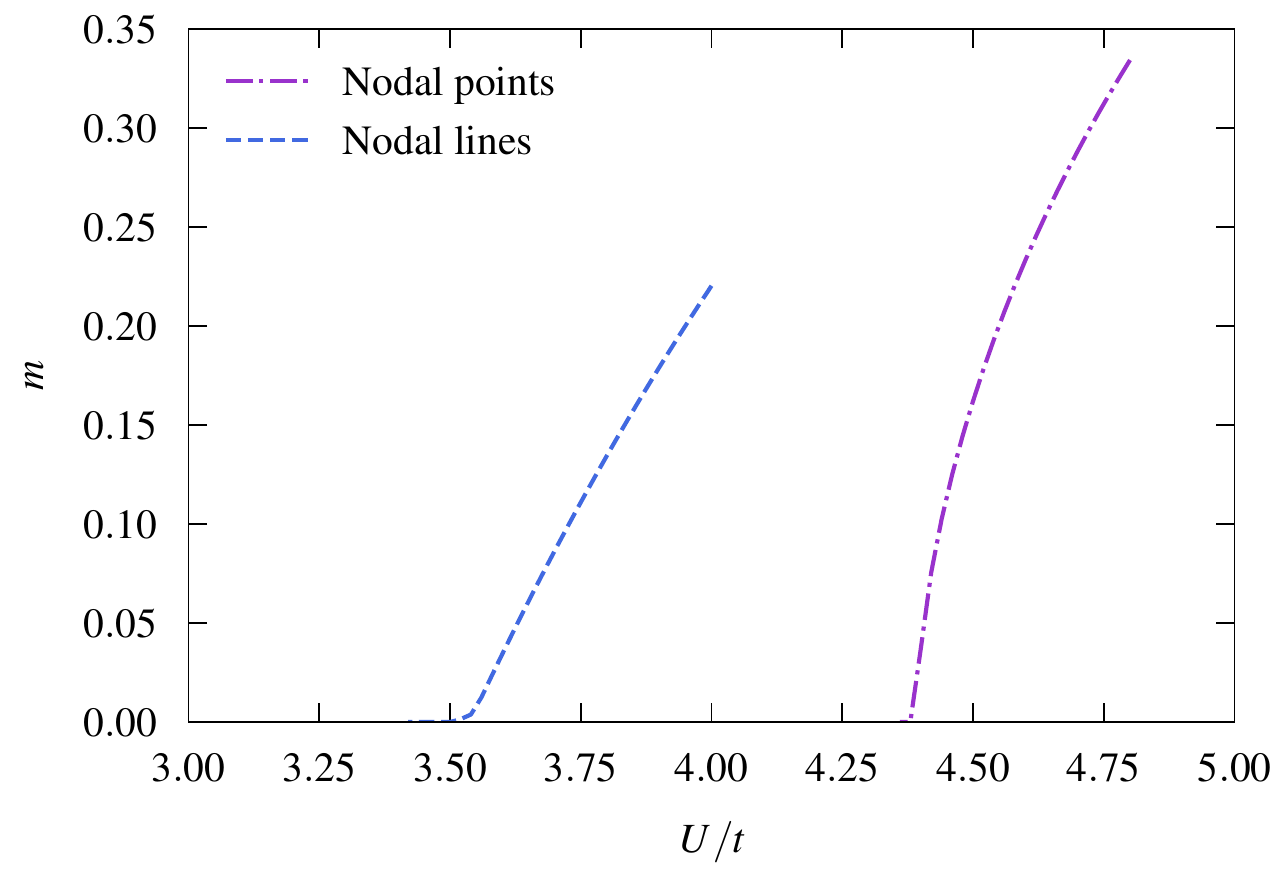} 
\caption{Mean-field AFM order parameter $m$ as a function of $U$ for the NP (purple dot-dashed line) and NL (blue dashed line) cases in the thermodynamic limit.}
\label{fig:MF}
\end{figure}

The distinction between the NP and NL cases is already visible at the mean-field level. By solving the self-consistent Eq.\,\eqref{eq:MF_order_parameter}, we obtain the order parameter as a function of $U$, as shown in Fig.\,\ref{fig:MF}. Both cases exhibit a finite critical interaction for the onset of AFM order: we find $U_{c,NP}/t \approx 4.392 $ and $U_{c,NL}/t \approx 3.538 $. That is, the nodal-line system becomes unstable toward AFM order at a smaller interaction strength than the nodal-point system. 

The behavior of the order parameter close to the transition also differs between the two cases. Assuming the thermodynamic limit for the NP case, the order parameter follows the conventional mean-field square-root behavior as a function of $U-U_c$, with the critical exponent $\tilde{\beta} \approx 1/2$ (not shown). On the other hand, for the NL case, it increases approximately linearly with $U-U_c$, i.e., with $\tilde{\beta} \approx 1$ (not shown). These differences for $U_c$ and $\tilde{\beta}$ are consistent with the distinct low-energy DOS of the two systems.
\footnote{Here, we denote the magnetization critical exponent by $\tilde{\beta}$ to avoid confusion with the inverse temperature $\beta=1/k_{\rm B}T$.}

Considering the NL limit and expanding the dispersion around $k_x=k_y=\pi/2$, with $\mathbf{q}_{\perp}=(q_x,q_y)$ being the momenta transverse to the nodal line, one obtains
\begin{equation}
E(\mathbf{q}_\perp, k_z, \Delta)
=
\pm 2t  \sqrt{ |\mathbf{q}_\perp|^2 + |\mathbf{q}_\perp|^4 \cos^2 (k_z) + \Delta^2}~.
\end{equation}
Here $k_z$ parametrizes the position along the nodal line. In the low-energy limit, $|\mathbf{q}_{\perp}|\to 0$, the term proportional to $|\mathbf{q}_{\perp}|^4$ is subleading, and the dispersion reduces to
\begin{equation}
E(\mathbf{q}_\perp, k_z, \Delta)
\approx
\pm 2t  \sqrt{ |\mathbf{q}_\perp|^2 + \Delta^2}~.
\end{equation}
Thus, the nonanalytic dependence of the mean-field free energy on $\Delta$ is governed by the two transverse momentum directions, while the integration along the nodal line contributes only a finite prefactor. This is the same scaling structure that appears in two-dimensional Dirac systems, such as the honeycomb lattice.
As a consequence, the NL case yields a mean-field magnetization exponent $\tilde{\beta}=1$, in contrast to the NP case, for which the conventional mean-field result $\tilde{\beta}=1/2$ is obtained.
Therefore, although the lattice is three-dimensional, the critical behavior of the NL semimetal is controlled by the codimension $d_c=2$ of the nodal manifold. That is, as the dispersion along the nodal line is subleading, we expect that it weakly modifies the leading scaling near the transition.

\subsection{Quantum Monte Carlo: equal-time results}

\begin{figure}[t]
\includegraphics[width=0.99\linewidth]{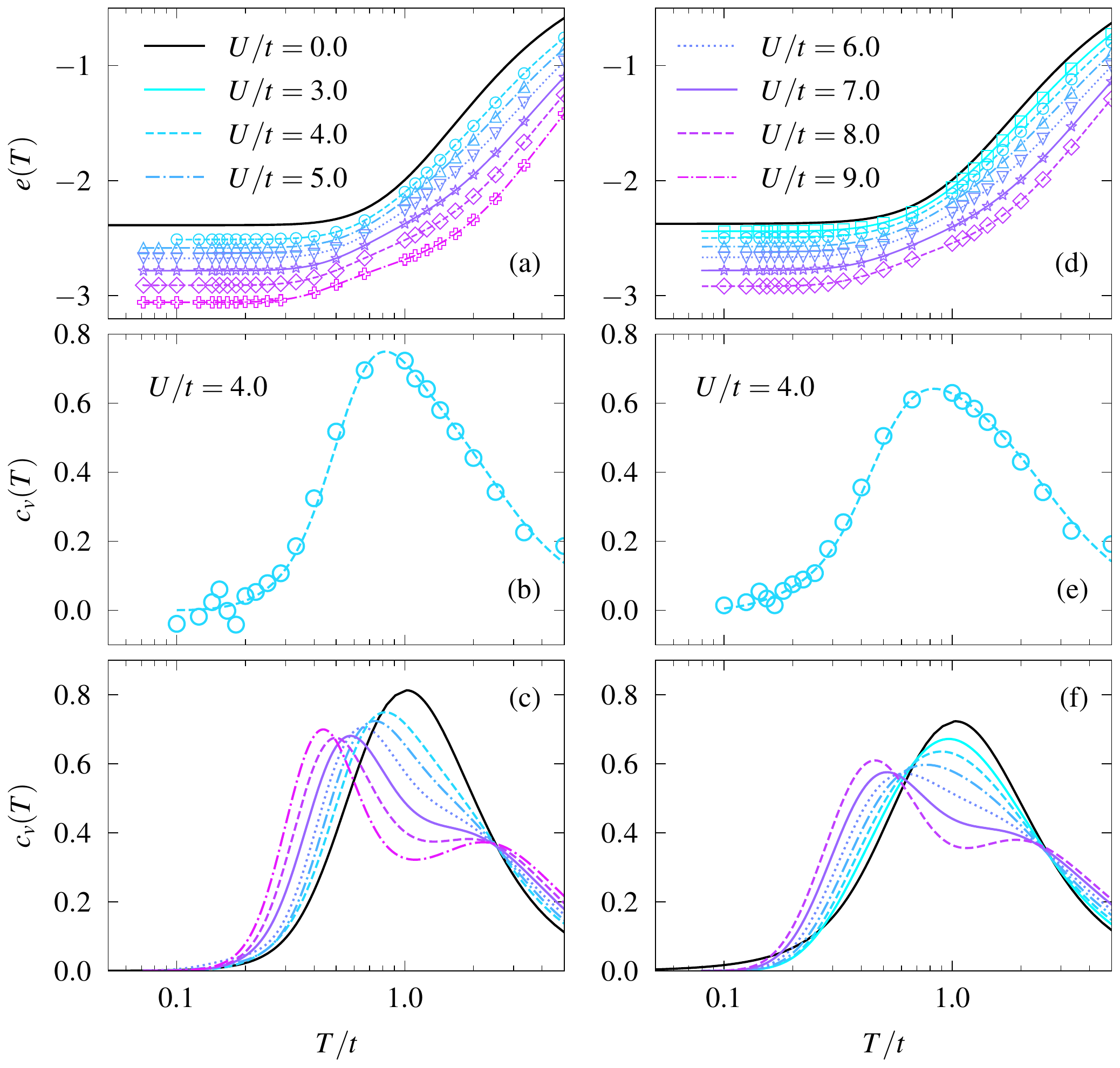} 
\caption{Energy density and specific heat for the NP and NL cases on a lattice with $N=4\times4^3$ sites. Panels (a) and (d) show $e(T)$, with symbols denoting DQMC data and lines the fits $e_{\rm fit}(T)$. Panels (b) and (e) compare the numerical derivative of $e(T)$ with the derivative of $e_{\rm fit}(T)$ for $U/t=4$. Panels (c) and (f) show the resulting $c_{\rm v}(T)$ for different interaction strengths from the fit.} 
\label{fig:specheat}
\end{figure}

We now turn to the interacting regime using DQMC simulations. Since the model is three-dimensional, AFM long-range order may occur at finite temperature. Therefore, as a first thermodynamic probe, we analyze the energy densities,
\begin{equation}
e(T)=\frac{\langle \mathcal{H}\rangle}{N},
\end{equation}
where $N=4 \times L^3$ is the total number of lattice sites, and $L$ is the linear size.
Figures~\ref{fig:specheat}\,(a) and \ref{fig:specheat}\,(d) display $e(T)$ for the NP and NL cases, respectively.
In fact, the quantity more directly related to a phase transition is the temperature derivative of the energy, namely the specific heat, $c_{\rm v}(T)=\frac{d e(T)}{dT}$.
As the numerical QMC data may be noisy, it is convenient to perform a nonlinear fit of the energy density as
\begin{equation}
e_{\rm fit}(T)
=
a_0
+
\sum_{n=1}^{M}
a_n e^{-\beta n\Delta},
\end{equation}
where $M$ is a cutoff typically chosen in the range $4$--$7$, while $a_n$ and $\Delta$ are fitting parameters~\footnote{The cutoff $M$ is chosen by comparing the numerical derivative of the energy data with the derivative obtained from $e_{\rm fit}(T)$.}.
This procedure provides a smooth estimate of the energy and avoids amplifying Monte Carlo fluctuations when taking temperature derivatives.

Figures~\ref{fig:specheat}\,(b) and \ref{fig:specheat}\,(e) show representative results for $U/t=4$ in the NP and NL cases, respectively. The symbols correspond to the numerical derivative of the raw QMC data, while the solid lines are obtained from $d e_{\rm fit}(T)/dT$, which agree well. Therefore, applying the same procedure to the remaining values of $U$, we obtain the specific heat shown in Figs.~\ref{fig:specheat}\,(c) and \ref{fig:specheat}\,(f) for the NP and NL cases, respectively.
For small interactions, both the NP and NL cases display a single broad maximum in $c_{\rm v}(T)$, resembling the noninteracting behavior. As $U/t$ increases, this structure evolves into a two-peak profile. The higher-temperature peak is naturally associated with single-particle excitations, while the lower-temperature feature is associated with low-lying collective spin-wave excitations, consistent with the onset of AFM order. In both cases, this low-temperature peak becomes more evident at large interaction strength, for instance at $U/t=9$.

\begin{figure}[t]
\centering
\includegraphics[width=0.85\linewidth]{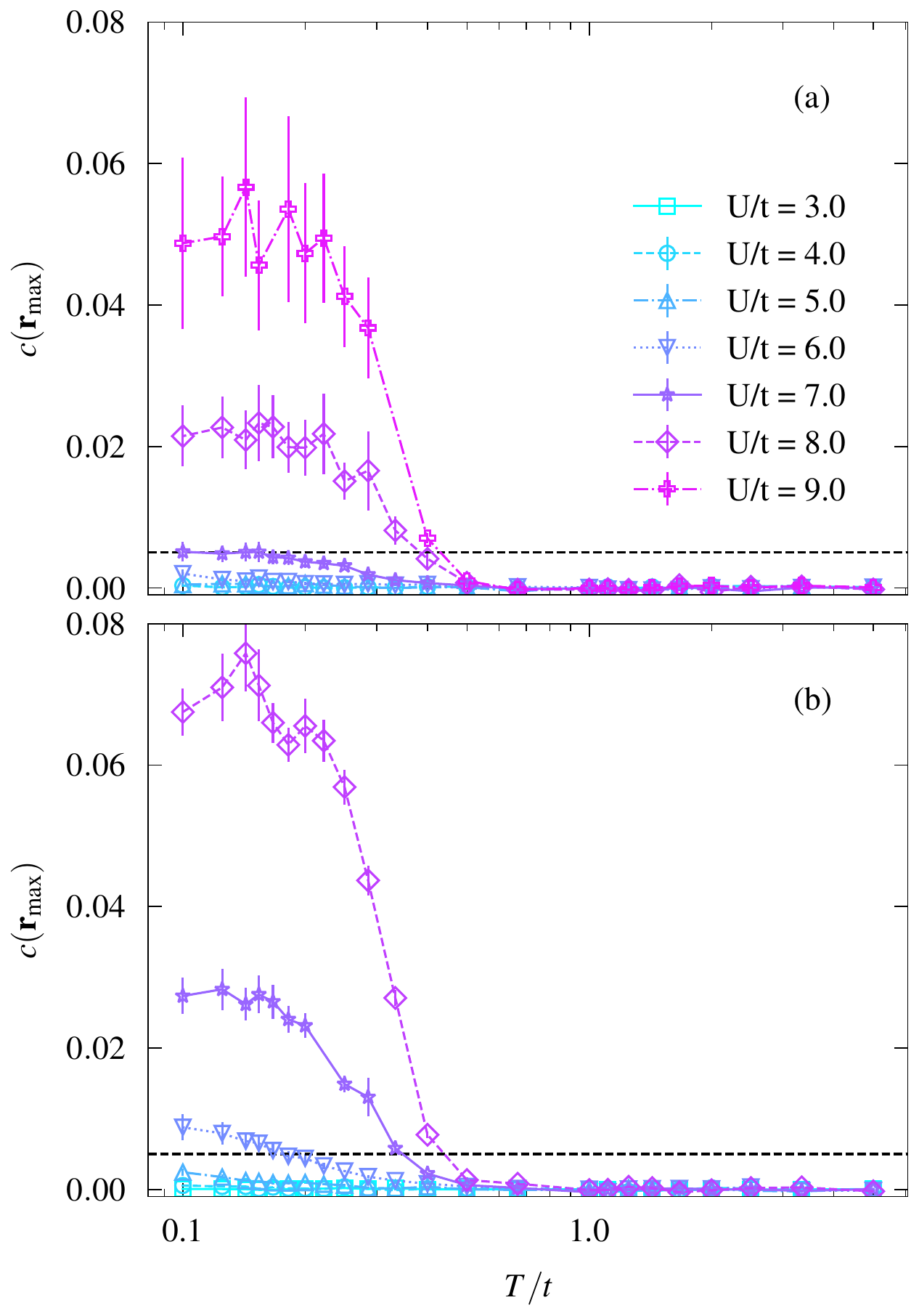} 
\caption{Long-distance same-orbital spin-spin correlation function as a function of temperature for (a) the NP case and (b) the NL case on a lattice with $L=4$ ($N=256$ sites).
The horizontal dashed line marks an arbitrary reference value below which the correlations are considered negligible in the present finite-size analysis.} 
\label{fig:spin-spin}
\end{figure}

Although the specific heat provides a clear indication of a phase transition as $U$ increases, it does not give a sharp determination of the quantum critical point. Therefore, to connect these thermodynamic signatures with magnetic correlations, we analyze the real-space spin-spin correlation function
\begin{equation}
c(\mathbf{r}_i-\mathbf{r}_j)
=
\langle S_i^z S_j^z\rangle ,
\end{equation}
as a function of temperature, with $S_i^z=n_{i\uparrow}-n_{i\downarrow}$. We focus on correlations between equivalent orbitals, averaging over the $A$-$A$, $B$-$B$, $C$-$C$, and $D$-$D$ contributions. In order to probe the longest-distance correlations available on the finite lattice, we take $ \mathbf{r}_{\rm max} \equiv  \mathbf{r}_i-\mathbf{r}_j=(L/2,L/2,L/2) $. Figure~\ref{fig:spin-spin} shows this quantity for $L=4$, corresponding to $N=256$ sites, for (a) the NP and (b) the NL cases. For comparison, we introduce a small reference value at $10^{-3}$, shown as the horizontal dashed line.
%Correlations below this scale are treated as negligible within the present finite-size analysis.

At high temperatures, the long-distance correlations are strongly suppressed for all interaction strengths, as expected from the short-range character of the spin correlations in this regime. As the temperature is lowered, however, $c(\mathbf{r}_{\rm max})$ increases significantly for sufficiently large $U/t$, indicating the onset of long-range magnetic correlations.
The temperature scale at which these long-distance correlations become relevant is consistent with the low-temperature peak observed in the specific heat, shown in Figs.~\ref{fig:specheat}\,(c) and (f), which is around $T/t\approx 0.4$ for large interactions. This supports the above interpretation that the low-temperature peak is associated with the development of a long-range AFM order.

This analysis also provides useful insights into the zero-temperature magnetic phase transition. As the temperature is lowered, the longest-distance correlations remain small for $U/t\lesssim7$ in the NP case, as shown in Fig.\,\ref{fig:spin-spin}\,(a), even at the lowest temperatures considered.
Similarly, in the NL case, shown in Fig.\,\ref{fig:spin-spin}\,(b), the correlations are strongly suppressed for $U/t\lesssim6$ and become appreciable only at larger values of $U/t$. Therefore, the onset of AFM ordering is expected to occur around these interaction scales, which explains the single-peak structure of the specific heat for weaker interactions.

\begin{figure}[t]
    \centering
    \includegraphics[width=0.95\linewidth]{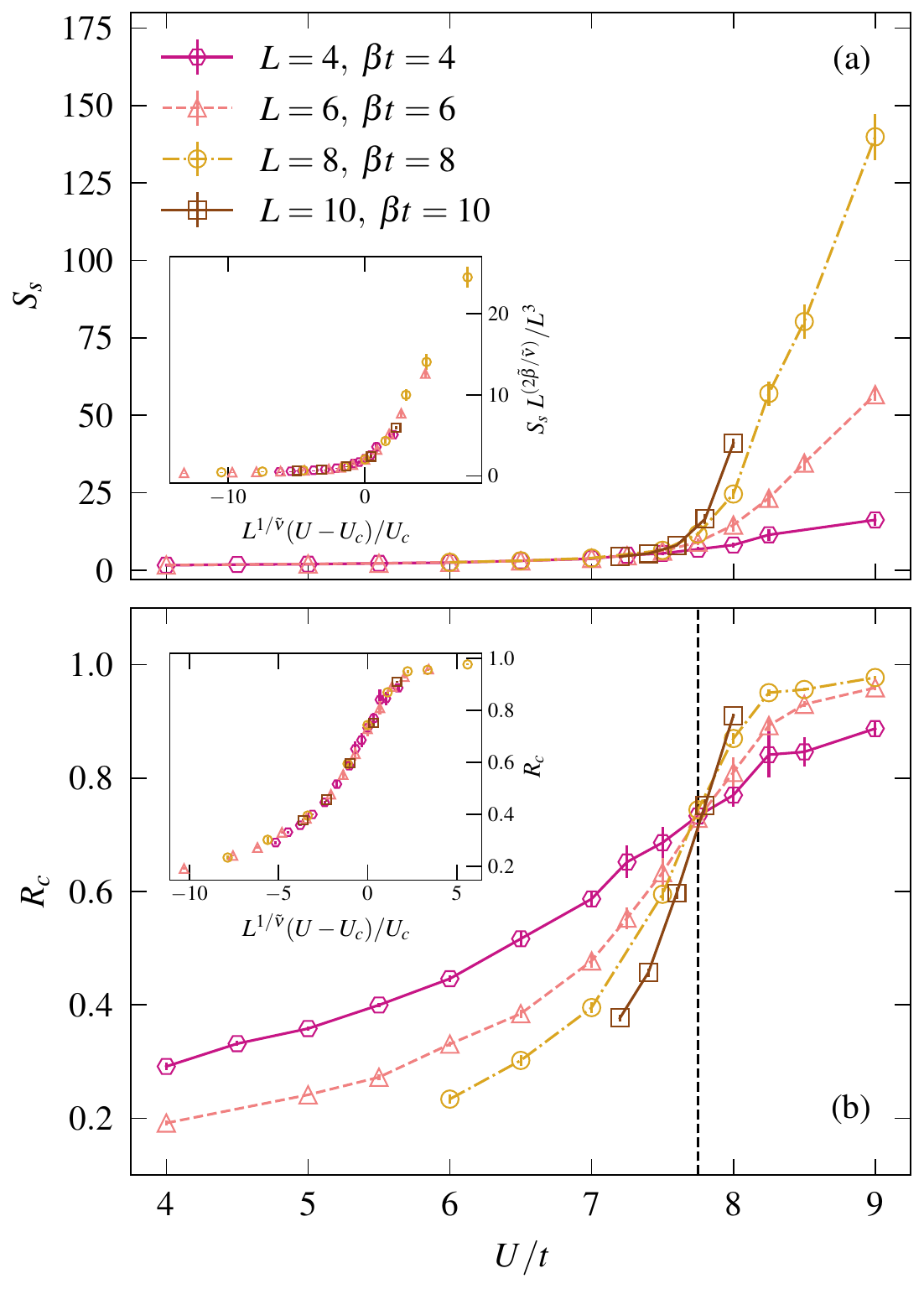}
    \caption{(a) Staggered AFM structure factor $S_s$ and (b) its corresponding correlation ratio $R_c$ as functions of $U/t$ for different system sizes in the NP case (cubic $\pi$-flux Hubbard model). The horizontal dashed line correspond to the critical point $U_{c}/t\approx 7.8$. The insets describe their respective finite-size scaling analysis, providing $\tilde{\nu}\approx\tilde{\beta} \approx 1/2$.}
    \label{fig:Saf_Rc_NP}
\end{figure}

Beyond these hints from the specific heat and $c(\mathbf{r}_{\rm max})$, the emergence of a long-range ordered phase at low $T$ is determined by a finite-size scaling analysis of the spin structure factor
\begin{equation}
S_s(\mathbf{q})
=
\frac{1}{N}
\sum_{i,j}
e^{i\mathbf{q}\cdot(\mathbf{r}_i-\mathbf{r}_j)}
\langle S_i^z S_j^z\rangle ,
\end{equation}
where $\mathbf{q}$ is a wavevector of the lattice. Because the unit cell is chosen as $2\times2\times1$, the staggered AFM order occurs at $\mathbf{Q}=(0,0,\pi)$, in the extended Brillouin zone. Hereafter, we denote $S_s\equiv S_s(\mathbf{Q})$ as the staggered AFM structure factor.\,\footnote{To account for the staggered AFM pattern within the $xy$ plane, we include an additional form factor $(-1)^{x_i+y_i}$ in the spin correlations. This phase factor, which alternates the sign between neighboring sites along the $x$ and $y$ directions, is used throughout but omitted from the notation above for brevity.}

Figure~\ref{fig:Saf_Rc_NP}\,(a) shows $S_s$ as a function of $U/t$ for the NP case, using lattices with $L=4$, $6$, $8$, and $10$. Due to the presence of Lorentz symmetry, we set $\beta = L$, in line with the low-energy theory that provides a dynamical exponent $z=1$. For weak interactions, $S_s$ remains small and only weakly dependent on system size, consistent with the stability of the semimetallic phase. At larger values of $U/t$, $S_s$ grows rapidly with $L$, signaling the development of long-range AFM correlations.
The change in the behavior occurs around $U/t \approx 7.5$, which provides a first estimate of the critical point.

%\begin{figure}[t]
%    \centering
%    \includegraphics[width=0.85\linewidth]{New_Figures/collapse_Rc_Ss_NP.pdf}
%    \caption{Finite-size scaling collapse for the NP case. Panel (a) shows the best collapse of the correlation ratio $R_c$, yielding $U_c/t=7.8$ and $\tilde{\nu}=1/2$. Panel (b) shows the corresponding collapse of the spin structure factor $S_s$, using the values of $U_c$ and $\tilde{\nu}$ obtained from panel (a), which yields $\tilde{\beta}=1/2$. }
%    \label{fig:Rc_collapse_NP_NL}
%\end{figure}

A more precise determination of $U_c$ is obtained by using the correlation ratio\,\cite{Kaul15,Sato2018,Otsuka2020}
\begin{equation}
R_c(L) = 1 - \frac{S_s(\mathbf{Q}-\delta\mathbf{q})}{S_s(\mathbf{Q})},
\label{eq:Rc}
\end{equation}
where $|\delta\mathbf{q}|=2\pi/L$ is the smallest momentum interval allowed on a lattice with linear size $L$. In the absence of long-range AFM order, the spin correlations remain short ranged and $R_c\rightarrow0$ in the thermodynamic limit. By contrast, in the ordered phase, $S_s(\mathbf{Q})$ grows more strongly than $S_s(\mathbf{Q}-\delta\mathbf{q})$, leading to $R_c\rightarrow1$. At the critical point, $R_c$ is a scale-invariant quantity.

Figure~\ref{fig:Saf_Rc_NP}\,(b) shows the correlation ratio as a function of the interaction strength for the corresponding data of panel (a)~\footnote{For the analysis of $R_{c}$, we chose $\delta\mathbf{q}=(0,0,2\pi/ L)$.}.
A clear crossing occurs near $U_c/t=7.8(1)$, signaling the onset of AFM long-range order for $U>U_{c}$, and determining the value of the critical point for the NP case.
This estimated critical strength is larger than the value reported in Ref.~\onlinecite{huang2020} for the same model. This difference may originate from details of the finite-size analysis, boundary conditions, and the range of lattice sizes considered. Since our simulations include lattices up to $N=4\times10^3$ sites with different boundary conditions, the present result provides a benchmark for the cubic $\pi$-flux Hubbard model.

We next examine the correlation-length exponent $\tilde{\nu}$ from the finite-size collapse of the correlation ratio. Since the crossing analysis gives $U_c/t \approx 7.8$, we use the scaling form
\begin{equation}
R_c(L,U) = \mathcal{F} \left[L^{1/\tilde{\nu}} \frac{(U-U_c)}{U_c} \right].
\end{equation}
As shown in the inset of Fig.\,\ref{fig:Saf_Rc_NP}\,(b), the data collapse is consistent with $\tilde{\nu}\approx 1/2$, in agreement with the mean-field exponent.

\begin{figure}[t]
    \centering
    \includegraphics[width=\linewidth]{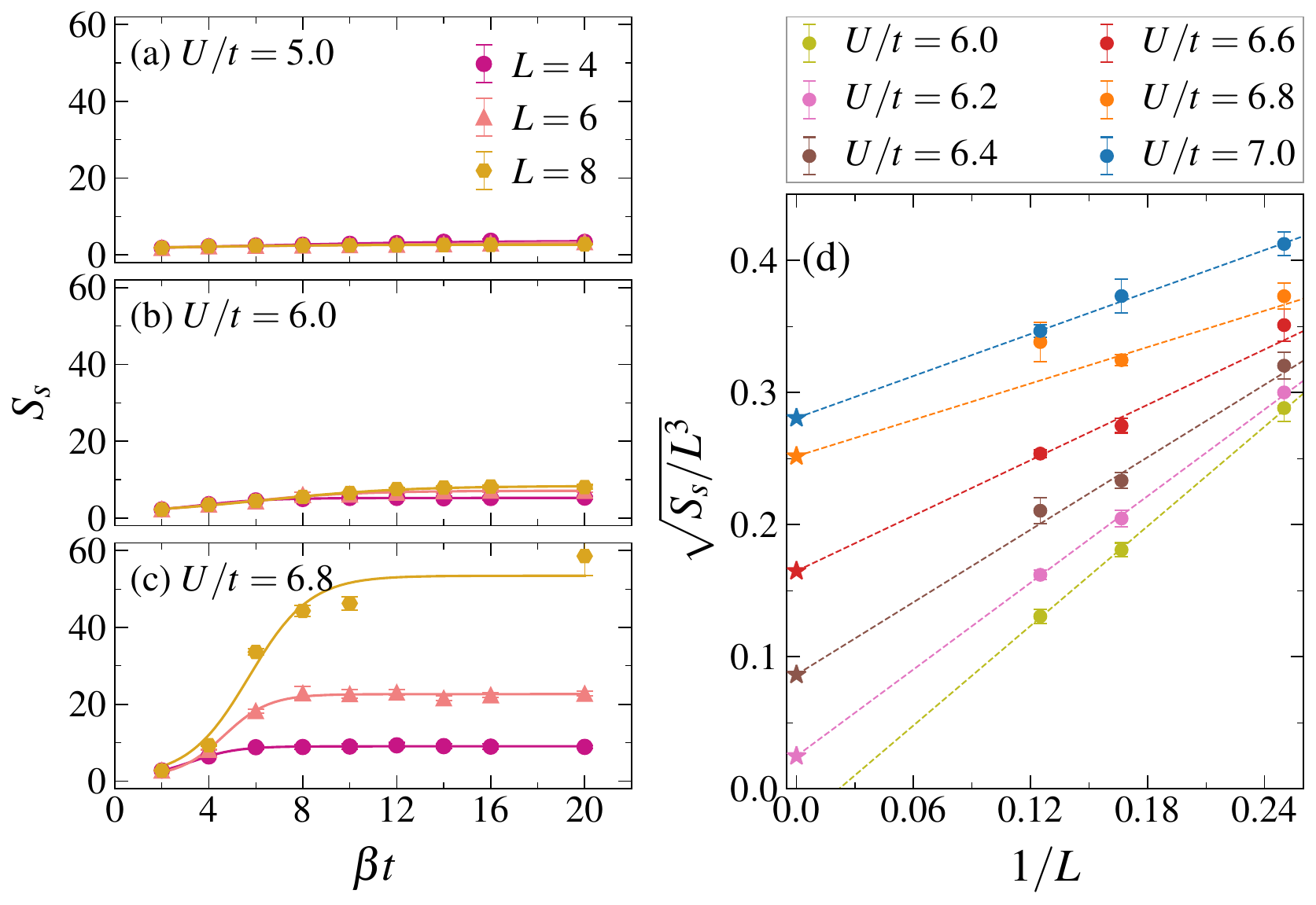}
    \caption{Staggered spin structure factor $S_s$ for the NL case as a function of inverse temperature for different system sizes at (a) $U/t=5$, (b) $U/t=6$, and (c) $U/t=6.8$. Panel (d) shows the low-temperature finite-size analysis of the corresponding order parameter $m(L)$.}
    
    \label{fig:Saf_vs_beta_NL}
\end{figure}

One may also estimate the order-parameter exponent $\tilde{\beta}$ from the scaling of $S_s(\mathbf{Q})$ \footnote{Here, we denote the magnetization critical exponent by $\tilde{\beta}$ to avoid confusion with the inverse temperature $\beta=1/k_{\rm B}T$.}, which obeys
\begin{equation}
S_s(L,U) = L^{d-2\tilde{\beta}/\tilde{\nu}} \mathcal{F} \left[ L^{1/\tilde{\nu}} \frac{(U-U_c)}{U_c} \right],
\end{equation}
with $d=3$. Using $U_c$ and $\tilde{\nu}$ obtained from the crossing of $R_c$, we can estimate the exponent $\tilde{\beta}$ from the data collapse of $S_s/L^{d-2\tilde{\beta}/\tilde{\nu}}$, as shown in the inset of Fig.\,\ref{fig:Saf_Rc_NP}\,(a).
We obtain $\tilde{\beta} \approx 1/2$, also consistent with the expected mean-field critical behavior of the NP transition, in line with the collapse of the correlation ratio.
Thus, these results provide an independent verification that the semimetal--AFM quantum phase transition in the cubic $\pi$-flux model is compatible with the $(3+1)$-dimensional Gross--Neveu mean-field scaling\,\cite{Moshe2003}.

\begin{figure}[t]
    \centering
    \includegraphics[width=0.85\linewidth]{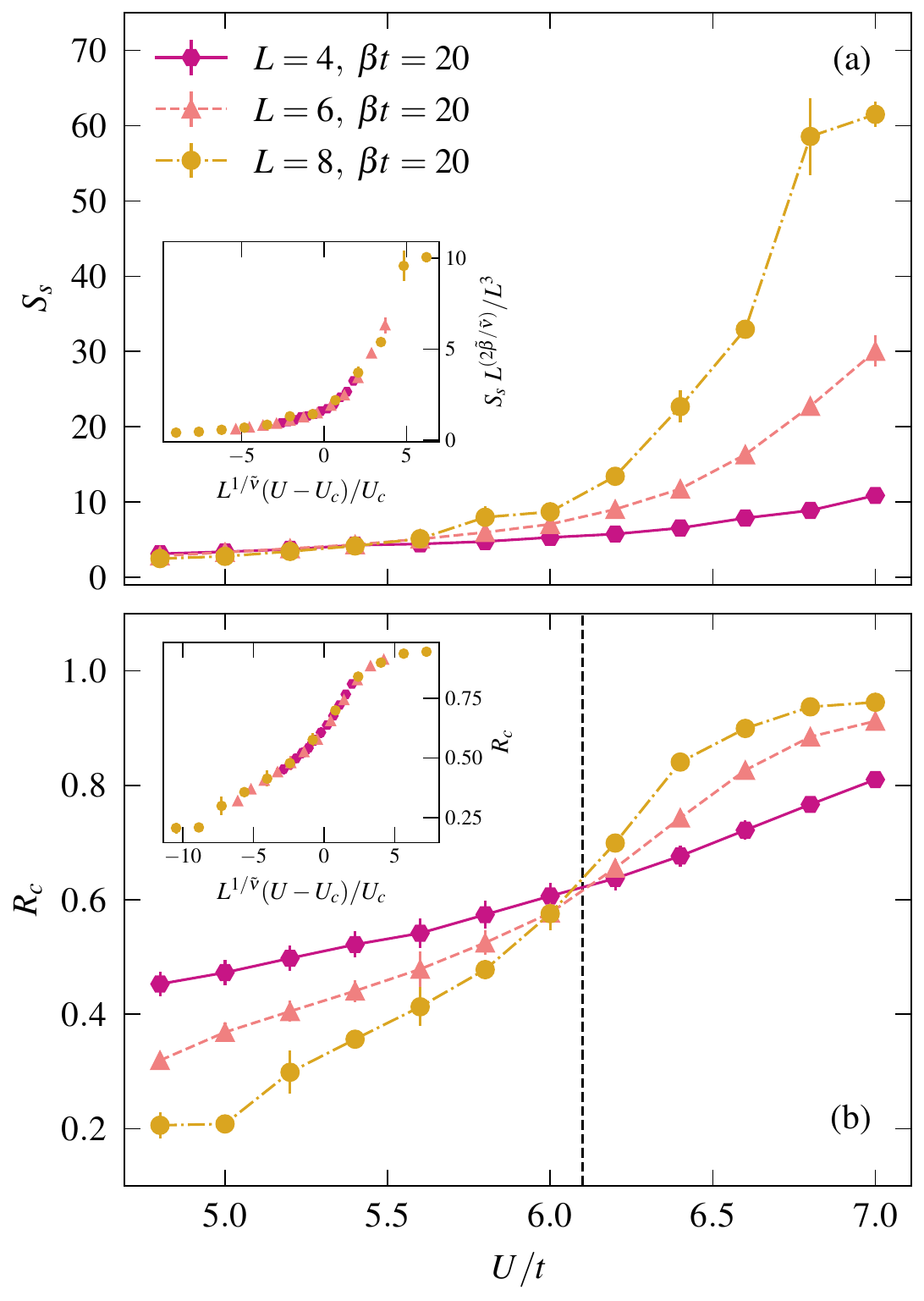}
    \caption{ (a) Staggered spin structure factor $S_s$ and (b) its correlation ratio $R_c$ for the NL case at fixed inverse temperature $\beta t=20$. The crossing of $R_c$ near $U_c/t\approx 6.1$ indicates the onset of AFM order. 
    The insets describe their respective finite-size scaling analysis, providing $\tilde{\nu}\approx\tilde{\beta} \approx 1/2$.
    }
    \label{fig:Saf_Rc_NL}
\end{figure}

We now turn to the NL case. In contrast to the NP limit, the nodal-line system is anisotropic and does not possess the same Dirac cone structure. Therefore, we do not impose the ground-state projection condition $\beta=L$ a priori. Although field-theoretical analyses suggest $z\approx 1$ for nodal-line semimetals \cite{jose2020}, here we adopt a more direct numerical procedure. We first analyze the temperature dependence of the spin structure factor and then perform the correlation ratio analysis only in a temperature regime where $S_s$ has saturated within the accessible system sizes.

Figure~\ref{fig:Saf_vs_beta_NL}\,(a)--(c) shows $S_s$ as a function of inverse temperature for several lattice sizes and interaction strengths. At high temperatures (small $\beta$), the spin correlations are short ranged and $S_s$ is weakly dependent on system size. As the temperature is lowered (larger $\beta$), the magnetic correlation length increases and may become comparable to the linear system size, depending on the interaction strength. 
For $U/t=5$, in panel (a), the spin structure factor remains weakly dependent on $L$ over the accessible temperature range, which is consistent with the absence of long-range AFM order. At $U/t=6$, a weak size dependence begins to develop, while for $U/t=6.8$ the growth with system size becomes more pronounced, as shown in panels (b) and (c), respectively.
The saturation at large $\beta$ indicates that the finite-size ground-state regime has been reached for the sizes considered.

 To further assess whether the low-temperature state is ordered, we perform a spin-wave finite-size scaling (FSS) analysis using
\begin{equation}
\sqrt{\frac{S_s(\mathbf{Q})}{L^3}} = m(L) + \frac{A}{L},
\end{equation}
where $m(L)$ is the AFM order parameter and $A$ is a fitting parameter. Figure~\ref{fig:Saf_vs_beta_NL}\,(d) shows this scaling analysis at $\beta=20$, where $S_s$ is already saturated for the sizes studied. The extrapolation gives a vanishing order parameter for $U/t=6$, whereas a finite intercept is obtained for $U/t=6.2$. Although these results should be interpreted with care, as the system size is still small, they indicate that the critical interaction in the NL case lies in the interval between $U/t\approx 6.0$ and $U/t\approx 6.2$.

This interval for the onset of AFM order is also visible in Fig.\,\ref{fig:Saf_Rc_NL}\,(a), which shows $S_s$ as a function of $U/t$ for different system sizes at fixed inverse temperature $\beta t=20$. Indeed, the growth of $S_s$ with $L$ becomes appreciable around $U/t\approx 6$, consistent with the finite-size extrapolation discussed above.
A more accurate estimate is obtained from the correlation ratio. Figure~\ref{fig:Saf_Rc_NL}\,(b) shows $R_c$ as a function of $U/t$ at $\beta t=20$. The curves for different lattice sizes cross near a common point, yielding our best estimate $U_c/t = 6.1(1)$  for the AFM transition in the NL case. This value is smaller than the critical coupling obtained for the NP system, consistent with the enhanced low-energy density of states associated with the nodal line.
A finite-size collapse of $R_c$ can also be performed, yielding an effective exponent close to $\tilde{\nu}\approx 1/2$. Performing a scaling of $S_s(\mathbf{Q})$, we obtain the magnetic critical exponent $\tilde{\beta}\approx 1/2$, as shown in the insets of Fig.~\ref{fig:Saf_Rc_NL}.

At this point, it is worth emphasizing that the mean-field analysis reveals that very large system sizes are required to recover the asymptotic scaling behavior associated with the nodal-line codimension $d_c=2$. Consequently, our QMC results for the NL case are expected to exhibit much stronger finite-size effects than those for the NP case, and quantitative estimates of the critical exponents should be interpreted with caution. We return to this issue below.

\subsection{Quantum Monte Carlo: unequal-time results}

We now complement the equal-time magnetic observables with dynamical information obtained from the unequal-time Green's function. While the spin structure factor and the correlation ratio diagnose the onset of AFM order, the single-particle spectral function allows us to follow how the low-energy fermionic excitations evolve across the transition.

\begin{figure}[t]
\centering
\includegraphics[width=1.0\linewidth]{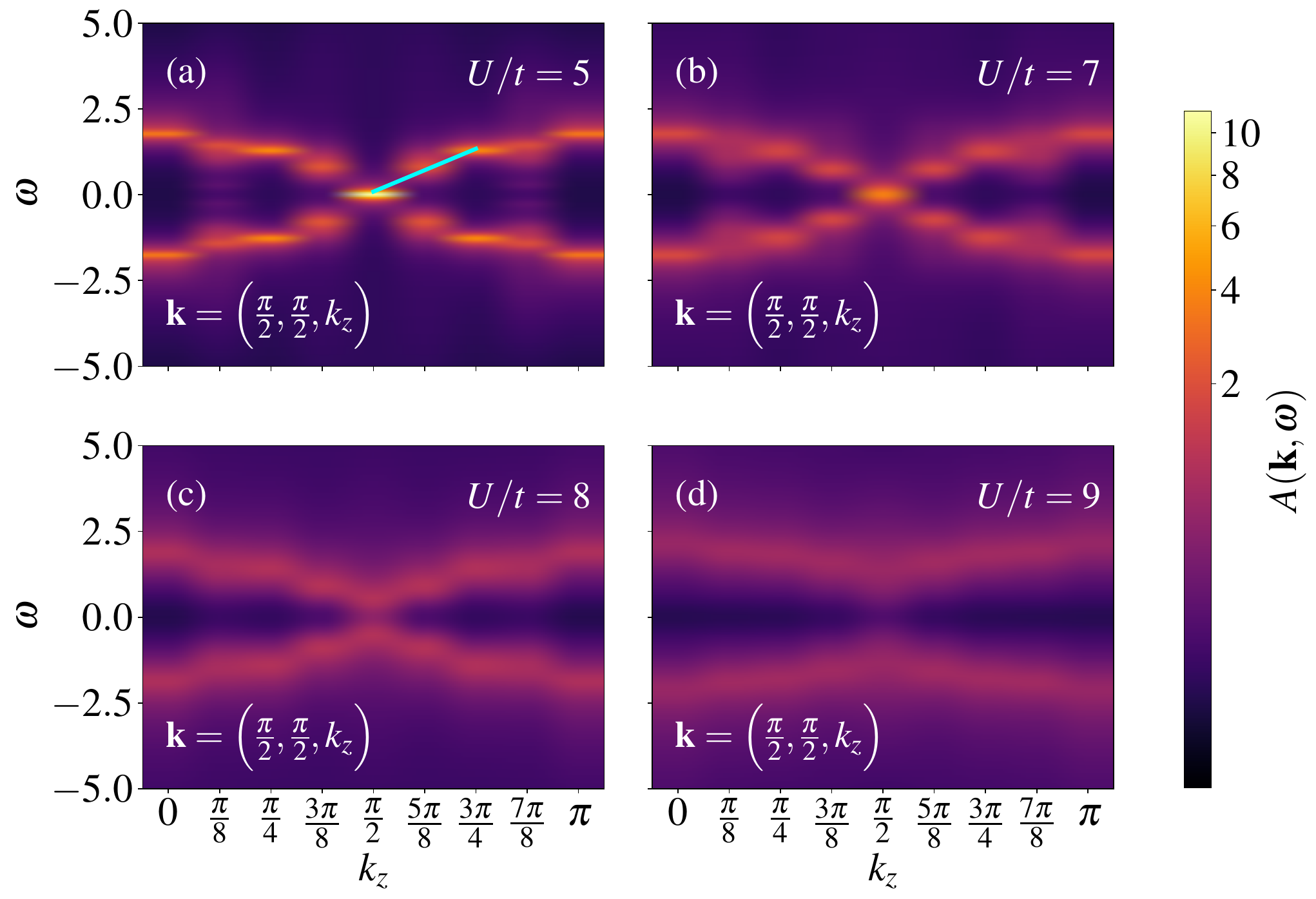} 
\caption{Momentum-resolved spectral function $A(\mathbf{k},\omega)$ for the nodal-point (cubic $\pi$-flux) case at fixed inverse temperature $\beta t=8$. The spectra are shown along the momentum line $\mathbf{k}=(\pi/2,\pi/2,k_z)$ for (a) $U/t=5$, (b) $U/t=7$, (c) $U/t=8$, and (d) $U/t=9$. The cyan line in panel (a) represents a linear fit to the peak positions near the nodal point, yielding an estimate $v_{\rm F}\approx 1.6\,ta$. } 
\label{fig:Akw_NP}
\end{figure}

In DQMC, the quantity measured directly is the imaginary-time Green's function, $G_{\alpha\alpha^{\prime}}(\mathbf{k},\tau)$, with $\alpha$ and $\alpha^{\prime}$ denoting orbital indices. Here, we consider the orbital-averaged Green's function
\begin{equation}
G(\mathbf{k},\tau)
=
\frac{1}{4}
\sum_{\alpha}
G_{\alpha\alpha}(\mathbf{k},\tau)~.
\end{equation}
Obtaining $A(\mathbf{k},\omega)$ from $G(\mathbf{k},\tau)$ requires an analytic continuation from imaginary time to real frequency,
\begin{equation}
G(\mathbf{k},\tau)
=
\int_{-\infty}^{\infty}
d\omega\,
K(\tau,\omega)
A(\mathbf{k},\omega),
\end{equation}
with $ K(\tau,\omega) = \frac{e^{-\tau\omega}} {1+e^{-\beta\omega}}$ being the fermionic kernel.
This inversion is numerically obtained by a stochastic analytic continuation (SAC) procedure\,\cite{sandvik1998,Beach2004}.
For both cases, we used a $4\times4\times8$ lattice, employing both PBC and APBC to access more wavevectors.
\footnote{It is worth noting that finite-size effects are partially mitigated by averaging results obtained with PBC and APBC. However, for momentum-resolved Green's functions, changing the boundary conditions changes the set of allowed wavevectors, so that such an average cannot generally be performed at fixed $\mathbf{k}$.}

Figure~\ref{fig:Akw_NP} shows the momentum-resolved spectral function $A(\mathbf{k},\omega)$ for the NP case, at fixed inverse temperature $\beta t=8$. 
Panels~\ref{fig:Akw_NP}\,(a)--\ref{fig:Akw_NP}\,(d) correspond to $U/t=5$, $7$, $8$, and $9$, respectively, with the spectra being shown along the line $\mathbf{k}=(\pi/2,\pi/2,k_z)$. 
For $U/t=5$ and $U/t=7$, the spectral weight clearly displays a Dirac-like dispersion centered around $k_z=\pi/2$, indicating that the semimetallic quasiparticles remain well defined in this interaction range. As $U/t$ is increased to $8$, the low-energy spectral weight near the nodal point is suppressed and a small gap begins to develop. Upon further increasing to $U/t=9$, the gap becomes much more pronounced, showing that the system has indeed entered an insulating regime. It is important to note that the opening of a small gap around $U/t\approx 8$ is consistent with the critical interaction $U_c/t\approx 7.8$ obtained from the equal-time magnetic correlations.

The spectral function also allows us to estimate the Fermi velocity in the semimetallic regime. To this end, we identify the peak positions of $A(\mathbf{k},\omega)$ for momenta close to the nodal point $(\pi/2,\pi/2,\pi/2)$ and fit the resulting dispersion linearly. This procedure is illustrated in Fig.\,\ref{fig:Akw_NP}\,(a), yielding $v_{\rm F}\approx 1.6\,ta$. This value is close to the noninteracting result, $v_{\rm F}=2\,ta$, indicating that the Hubbard interaction only weakly renormalizes the Fermi velocity before the onset of the insulating phase. 

\begin{figure}
    \centering
    \includegraphics[width=0.9\linewidth]{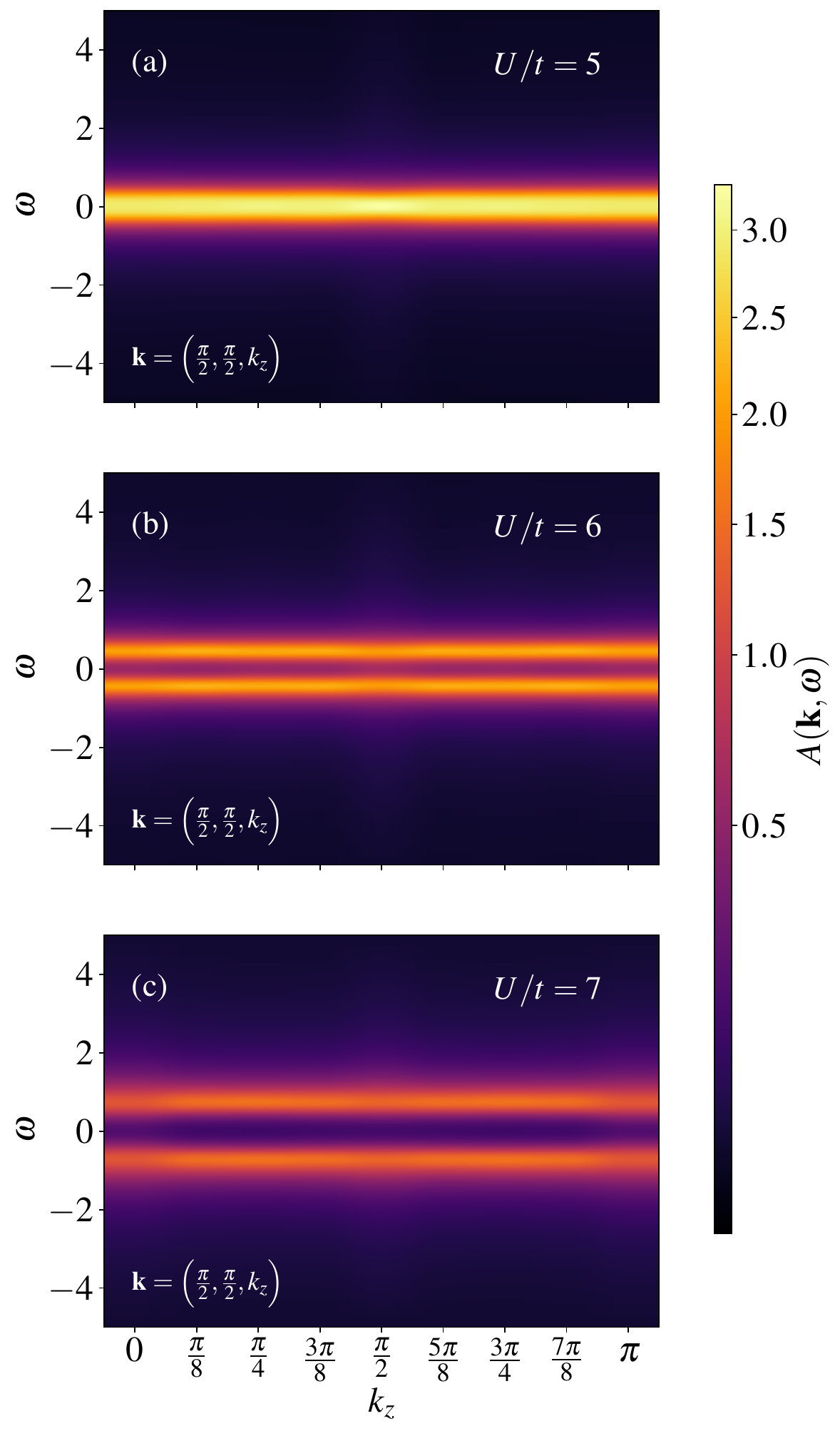}
    \caption{Momentum-resolved spectral function $A(\mathbf{k},\omega)$ for the nodal-line case along $\mathbf{k}=(\pi/2,\pi/2,k_z)$ for (a) $U/t=5$, (b) $U/t=6$, and (c) $U/t=7$. Here, we fix the inverse temperature at $\beta t=8$ on a $4\times4\times8$ lattice.}
    \label{fig:GreenF_NL}
\end{figure}

We now turn to the NL case, where a similar gap opening is observed around $U/t\approx 6$. Figure~\ref{fig:GreenF_NL} shows $A(\mathbf{k},\omega)$ along the nodal-line direction, $\mathbf{k}=(\pi/2,\pi/2,k_z)$, for (a) $U/t=5$, (b) $U/t=6$, and (c) $U/t=7$, at fixed $\beta t=8$. For $U/t=5$, the low-energy spectral weight exhibits no clear distortion or appreciable momentum dependence along the nodal-line direction, consistent with the persistence of a gapless nodal-line semimetal. Increasing the interaction strength to $U/t=6$ leads to the emergence of a small gap in the spectral function, consistent with the critical interaction $U_c/t=6.1(1)$ estimated from the crossing of the correlation ratio. For $U/t=7$, the gap becomes more pronounced, indicating the development of an insulating phase. Thus, the spectral analysis provides further evidence of the interaction-driven semimetal-to-insulator transition in the NL case, in agreement with the equal-time magnetic correlations.

\begin{figure}
    \centering
    \includegraphics[width=0.9\linewidth]{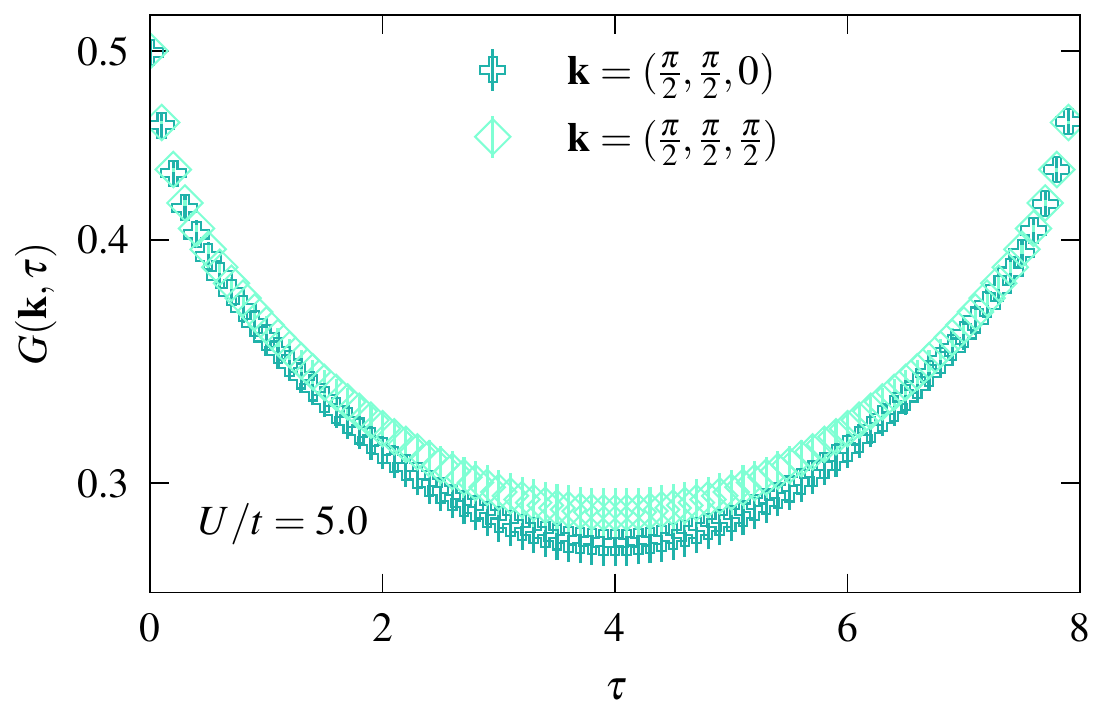}
    \caption{Unequal-time Green's functions as a function of imaginary time $\tau$ for $\mathbf{k}=(\pi/2,\pi/2,0)$ and $\mathbf{k}=(\pi/2,\pi/2,\pi/2)$ in the NL case, at fixed $U/t=5$ and $\beta t=8$.}
    \label{fig:GreensF}
\end{figure}

Before proceeding, it is worth examining the unequal-time Green's functions directly, particularly at $\mathbf{k}=(\pi/2,\pi/2,0)$ and $\mathbf{k}=(\pi/2,\pi/2,\pi/2)$, as shown in Fig.\,\ref{fig:GreensF} for $\beta t=8$. This comparison allows us to investigate possible differences between these two momenta without relying on SAC. Interestingly, we observe a small but noticeable difference (beyond error bars) between the two Green's functions, with the correlations at $k_z=0$ being more suppressed than those at $k_z=\pi/2$ at $\tau = \beta / 2$. This suggests that interactions may introduce a weak momentum dependence along the nodal line, which is not resolved in the spectral functions obtained through SAC. Therefore, although our results indicate that the nodal-line structure remains largely robust against interactions, a definitive assessment of possible small interaction-induced distortions requires complementary methods with higher spectral resolution.

\subsection{Weak-coupling renormalization-group analysis}
\label{sec:RG}

We now perform a weak-coupling renormalization-group analysis of the interacting nodal-line semimetal phase. The details of the methodology are presented in Appendix~\ref{app:RG}.
Close to the nodal line, the noninteracting Hamiltonian takes the long-wavelength form
\begin{equation}
\hat H_{\rm NL}^{(0)}\approx -v(q_x\Gamma_1+q_y\Gamma_2)+\kappa q_\perp^2\cos k_z\,\Gamma_3,
\label{eq:RG_H0_main}
\end{equation}
where $v=\kappa=2t$ and $q_\perp^2=q_x^2+q_y^2$. To leading order in momentum the system may therefore be viewed as a continuous stack of $(2+1)$-dimensional Dirac cones labeled by $k_z$.
As discussed above, this dimensional reduction of the low-energy dispersion leads to the same mean-field critical exponents as those of the two-dimensional honeycomb lattice.

The Wilsonian transformation integrates a tubular shell in the transverse momenta, $\Lambda/b<q_\perp<\Lambda$, while leaving $k_z$ unscaled. Here $b>1$ is the coarse-graining factor, and we introduce the logarithmic RG scale
$\ell\equiv\ln b$, such that increasing $\ell$ corresponds to flowing toward the infrared. Accordingly, the running transverse cutoff is $\Lambda(\ell)=\Lambda_0e^{-\ell}$ and the associated energy scale is
$E_\Lambda(\ell)=v\Lambda(\ell)$. Tree-level power counting then gives $[U]=-1$,  and $[\Delta]=+1$,
where the latter denotes the coefficient of the symmetry-allowed quadratic operator
\begin{equation}
\delta\hat H=\Delta\cos k_z\,\Gamma_3.
\label{eq:generated_operator_main}
\end{equation}
Thus, although the short-range Hubbard interaction is irrelevant at the Gaussian nodal-line fixed point, the quadratic deformation is a relevant perturbation.

At half filling, the first-order Hartree contribution vanishes in the particle-hole-symmetric formulation. The leading momentum-dependent correction therefore arises from the second-order self-energy. In compact notation, the corresponding diagram gives
\begin{equation}
\Sigma_{\sigma}^{(2)}(k)=U^2\int_q\Pi_{\bar{\sigma}}^{(0)}(q)\,G_{\sigma}^{(0)}(k-q),
\label{eq:selfenergy_second_order}
\end{equation}
where $k=(\mathbf{k},i\omega_n)$, $\int_q\equiv T\sum_{\Omega_m}\int_{\mathbf q}$, and
\begin{equation}
\Pi_{\bar{\sigma}}^{(0)}(q)=-\int_pG_{\bar{\sigma}}^{(0)}(p)G_{\bar{\sigma}}^{(0)}(p+q)
\label{eq:ph_bubble}
\end{equation}
is the bare particle-hole bubble of the opposite-spin fermions. The Green's functions in Eqs.\,\eqref{eq:selfenergy_second_order} and \eqref{eq:ph_bubble} retain the orbital matrix structure of the
lattice Hamiltonian; the explicit evaluation of the loop integrals is given in Appendix~\ref{app:RG}.

The contribution responsible for reconstructing the nodal manifold is obtained by projecting the static self-energy onto the $\Gamma_3$ channel. Along the original nodal line, $\mathbf{k}=(\pi/2,\pi/2,k_z)$, the leading harmonic is
\begin{equation}
\Sigma_3^{(2)}(k_z,0)=\Delta_U\cos k_z+\cdots,
\label{eq:selfenergy_main}
\end{equation}
with $\Delta_U=c_3\frac{U^2}{t}$, and $c_3\approx 2.91\times10^{-3}$. Thus, although the onsite Hubbard interaction is irrelevant at the
Gaussian nodal-line fixed point, it generates at $O(U^2)$ the symmetry-allowed single-particle perturbation
$\Delta_U\cos k_z\,\Gamma_3$, which is relevant under the low-energy scaling. Importantly, this reconstruction occurs without an accompanying weak-coupling instability toward a symmetry-broken many-body state: the Hubbard interaction itself remains perturbatively irrelevant, while its second-order self-energy reshapes the low-energy band structure.

\begin{figure}[t]
\centering 
\includegraphics[width=\linewidth]{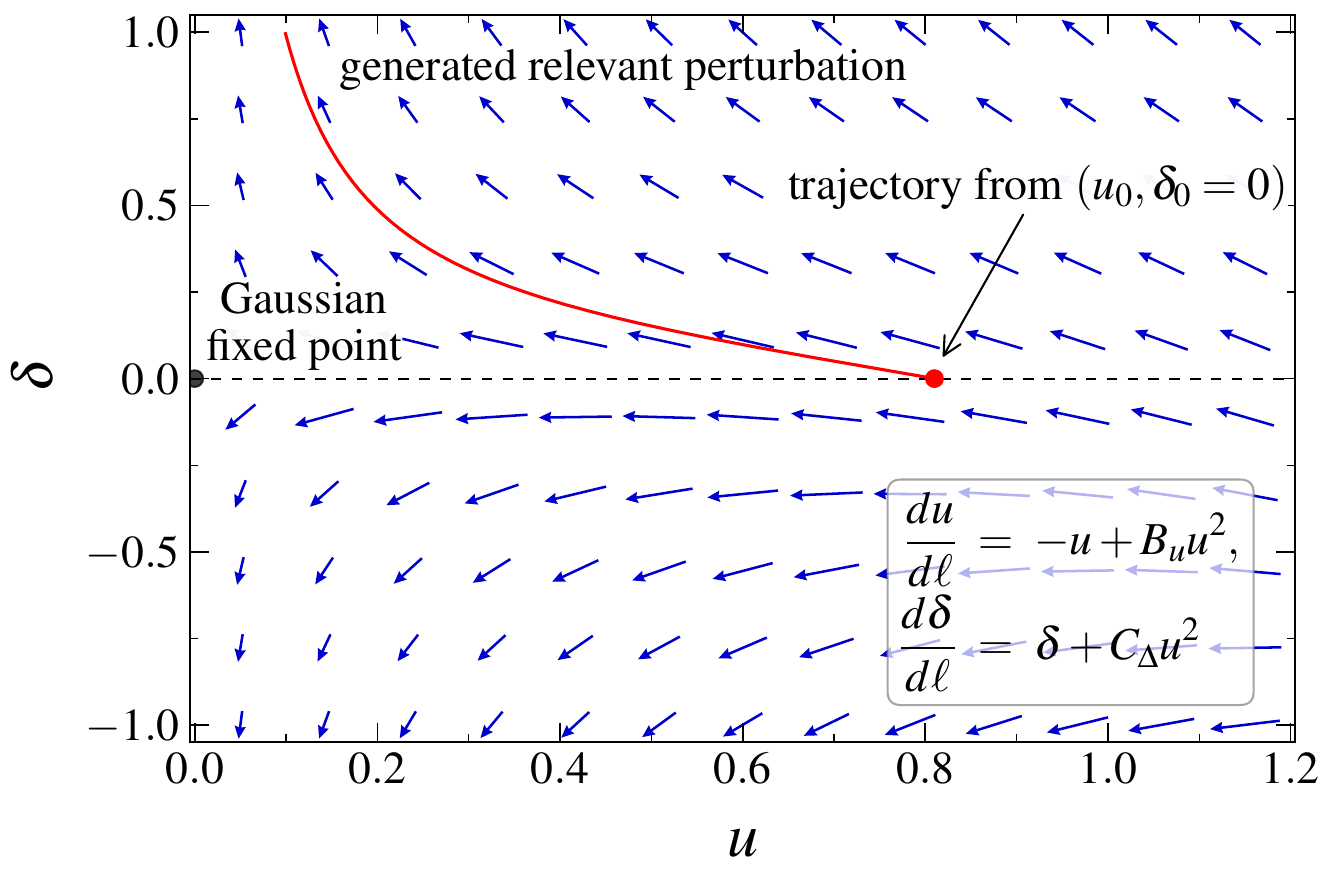} 
\caption{Weak-coupling RG flow of the dimensionless Hubbard interaction $u$ and the quadratic perturbation $\delta=\Delta/E_\Lambda$. The Gaussian fixed point at $(u,\delta)=(0,0)$ and trajectories starting from the microscopic condition $\delta_0=0$ are shown.}
\label{fig:RG_Flow}
\end{figure}

Introducing the dimensionless Hubbard coupling $u$ and $\delta=\Delta/E_\Lambda$, with $E_\Lambda=v\Lambda$, the corresponding weak-coupling RG equations are
\begin{align}
\frac{du}{d\ell}
&=
-u+B_u u^2+\cdots,
\label{eq:RG_u_main}
\\
\frac{d\delta}{d\ell}
&=
\delta+C_\Delta u^2+\cdots.
\label{eq:RG_delta_main}
\end{align}
The first equation expresses the perturbative irrelevance of the short-range interaction, whereas the $C_\Delta u^2$ term in the second equation is precisely the source term associated with the interaction-generated self-energy in Eq.\,\eqref{eq:selfenergy_main}.

For $u_0\ll1$, the nonlinear correction to the interaction flow is subleading, so that
\begin{equation}
u(\ell)=u_0e^{-\ell}.
\label{eq:u_solution_main}
\end{equation}
Starting from the microscopic nodal-line Hamiltonian, for which $\delta_0=0$, one then obtains
\begin{equation} 
\delta(\ell) = \frac{C_\Delta u_0^2}{3} \left( e^\ell-e^{-2\ell} \right).
\label{eq:delta_solution_main}
\end{equation}
The corresponding weak-coupling trajectories in the $(u,\delta)$ plane are illustrated in Fig.\,\ref{fig:RG_Flow}, for fixed $q_0=0.35$ and $v=2t$.
Hence the interaction itself flows to zero, while the relevant single-particle operator generated by it grows under coarse graining. Since the running cutoff scales as $E_\Lambda\propto e^{-\ell}$, the corresponding physical amplitude remains finite in the infrared. Matching the continuum flow to the lattice self-energy gives $\Delta_{\rm phys}=c_3 U^2/t$.

This perturbation qualitatively changes the topology of the zero-energy manifold. Along the original nodal line,
\begin{equation}
\hat H_{\rm eff}\left(\frac{\pi}{2},\frac{\pi}{2},k_z\right)=\Delta_{\rm phys}\cos k_z\,\Gamma_3,
\end{equation}
so that zero energy survives only for
$k_z=\pm\frac{\pi}{2}$.
The weakly interacting infrared theory is consequently not a nodal-line semimetal but a strongly anisotropic nodal-point semimetal with longitudinal velocity
$v_z\approx c_3 U^2 / t \ll 2t$.
However, the corresponding induced gap is small,
which explains why a broad intermediate-energy regime may appear nodal-line-like in finite-size simulations.

\begin{figure}[t]
\centering
\includegraphics[width=\linewidth]{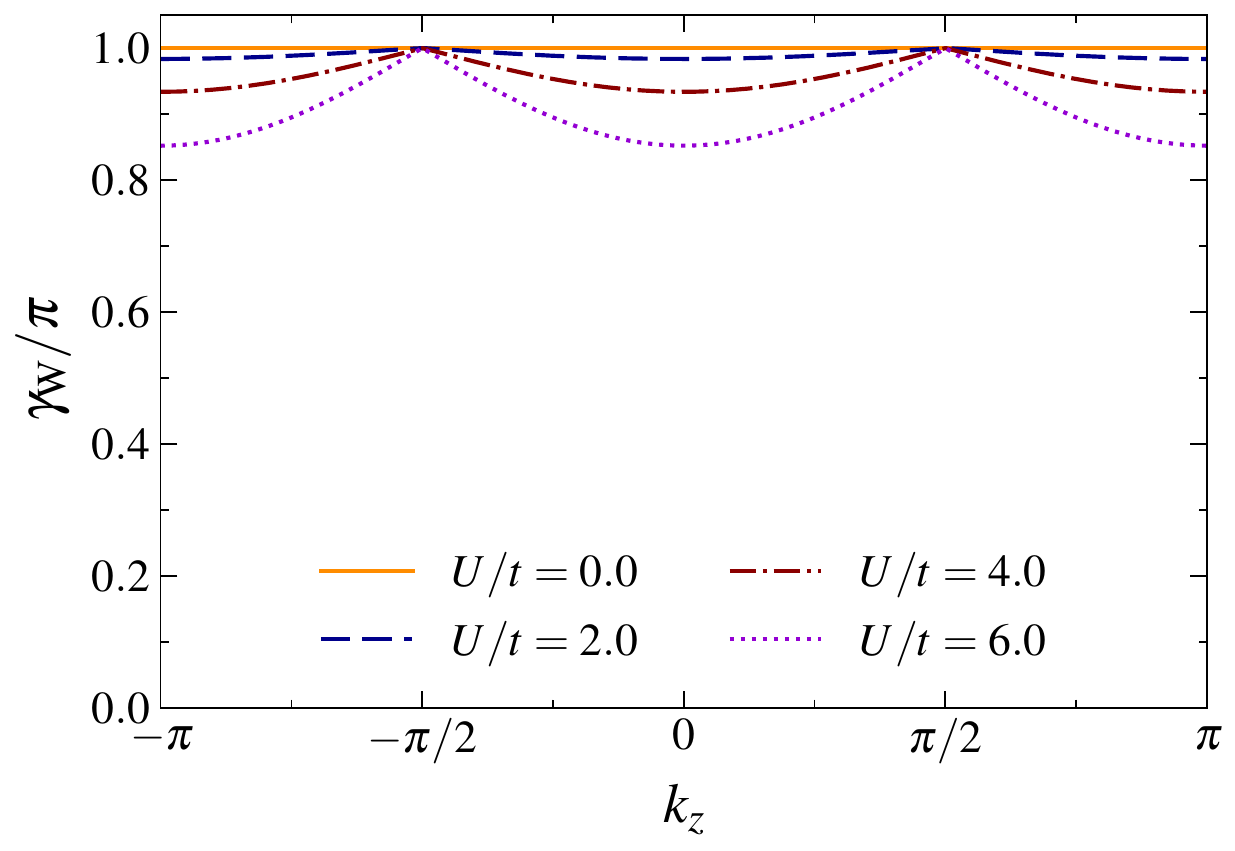}
\caption{Principal Wilson-loop phase $\gamma_{\rm W}$ evaluated along the original nodal line as a function of $k_z$ for representative interaction strengths. The interaction-induced scale is parametrized as $\Delta_{\rm phys}=c_3U^2/t$.}
\label{fig:Wilson_Loop}
\end{figure}

\subsection{Berry phase and Wilson-loop evolution}
\label{sec:Berry}

The interaction-driven reconstruction of the nodal manifold can also be characterized geometrically. In the noninteracting limit, a closed loop in the $(q_x,q_y)$ plane that links the nodal line encloses a Dirac singularity and carries a Berry phase $\pi$ modulo $2\pi$, as expected for a band-contact line \cite{MikitikSharlai1999,MikitikSharlai2008}. In the present four-band model, however, the occupied orbital subspace is twofold degenerate. The corresponding gauge-invariant quantity is therefore the Wilson loop, whose eigenvalues may be written as $e^{i\gamma_{\rm W}}$. In the decoupled Dirac sectors, the Wilson-loop eigenphase coincides with the corresponding Berry phase modulo $2\pi$. Hence, in the noninteracting nodal-line limit, both occupied sectors have eigenvalue $-1$, corresponding to
$\gamma_{\rm W}=\pi$ along the entire nodal line.

Once the interaction-generated term $\Delta_{\rm phys}\cos k_z\,\Gamma_3$ is included, a generic fixed-$k_z$ slice acquires an effective Dirac mass $m(k_z)=\Delta_{\rm phys}\cos k_z$. As outlined in Appendix \ref{app:berry}, for a circular transverse loop of radius $q_0$, the principal Wilson-loop phase can be written as
\begin{equation}
\gamma_{\rm W}(k_z;q_0)=\pi\left[1-\frac{|\Delta_{\rm phys}\cos k_z|}
{\sqrt{v^2q_0^2+\Delta_{\rm phys}^2\cos^2k_z}}\right].
\label{eq:Berry_kz_main}
\end{equation}
This expression is restricted to the principal branch $0\leq\gamma_{\rm W}\leq\pi$. In the noninteracting limit,
$\Delta_{\rm phys}\rightarrow0$, one recovers $\gamma_{\rm W}=\pi$ for every $k_z$. For finite interaction, the phase departs from $\pi$ on the gapped fixed-$k_z$ slices, whereas it remains pinned to $\gamma_{\rm W}=\pi$ at
$k_z=\pm\pi/2$, where the interaction-generated mass vanishes. In the limit of a shrinking transverse loop, $q_0\rightarrow0$, one furthermore obtains $\gamma_{\rm W}\rightarrow0$ for every gapped slice, while the value $\pi$ survives at the two nodal points.

Figure~\ref{fig:Wilson_Loop} illustrates this evolution for representative interaction strengths, using
$\Delta_{\rm phys}=c_3U^2/t$. The noninteracting system exhibits a uniform $\pi$ phase along the complete nodal line. Increasing $U$ progressively suppresses the Wilson-loop phase away from $k_z=\pm\pi/2$, while leaving the two surviving nodal points pinned at $\pi$. The evolution of $\gamma_{\rm W}$ therefore provides a direct geometrical signature of the interaction-driven reconstruction of the nodal line into isolated nodal points.

%%%%%%%%%%%%%%%%%%%%%%%%%%%%%%%%%%%%%%%%%%%%%%%%%%
\section{Conclusions}
\label{sec:conclusions}
%%%%%%%%%%%%%%%%%%%%%%%%%%%%%%%%%%%%%%%%%%%%%%%%%%

In this work, we investigated the effects of local electronic
interactions in three-dimensional semimetals with Dirac nodal-point
and Dirac nodal-line Fermi surfaces. By considering two closely
related lattice models, we directly examined how
the dimensionality of the nodal manifold affects the stability
of the semimetallic phase and the interaction-driven transition
toward antiferromagnetic order. To this end, we combined
mean-field calculations, determinant quantum Monte Carlo
simulations, and a weak-coupling renormalization-group analysis.

Already at the noninteracting and mean-field levels, the two
systems display important differences. While the density of
states vanishes quadratically in energy for the nodal-point
case, it vanishes linearly for the nodal-line system. As a
consequence, the latter is more susceptible to electronic interactions and it is expected to develop antiferromagnetic order at a smaller critical coupling. Within mean-field theory, we obtain $U_{c,\mathrm{NP}}/t \approx 4.39$ and $U_{c,\mathrm{NL}}/t \approx 3.54$, for the NP and NL cases, respectively.
In addition, the mean-field order parameter displays distinct critical behavior in the two cases. The nodal-point system exhibits the conventional mean-field behavior $\widetilde{\beta}=1/2$, whereas the nodal-line system approaches $\widetilde{\beta}=1$ in the thermodynamic limit.
These features (the codimension-two low-energy scaling, including the linear density of states and the associated mean-field critical behavior) are expected to apply broadly to smooth nodal lines, in particular those with linear transverse dispersion. 

The unbiased DQMC simulations confirm that both systems remain stable against weak-coupling antiferromagnetic ordering and undergo a transition to an antiferromagnetic insulating state only above a finite interaction strength. For the nodal-point model, our finite-size analysis gives $U_c/t=7.8(1)$, and the scaling of the correlation ratio and spin structure factor is compatible with $\widetilde{\nu}=1/2$ and $\widetilde{\beta}=1/2$.
Therefore, we provide an independent verification that the quantum phase transition in the cubic $\pi$-flux Hubbard model is compatible with the $(3+1)$-dimensional Gross--Neveu mean-field scaling\,\cite{Moshe2003}.
The single-particle spectral function further shows that the
Dirac-like quasiparticles survive throughout the semimetallic regime, with only a moderate renormalization of the Fermi velocity, while a gap develops upon entering the antiferromagnetic phase.

For the nodal-line system, the transition occurs at the
smaller interaction strength $U_c/t=6.1(1)$. This reduction
of the critical coupling is consistent with the enhanced
low-energy density of states associated with the extended
nodal manifold. The thermodynamic and magnetic observables
provide a consistent picture of the transition, with the
low-temperature structure in the specific heat accompanying
the development of long-range spin correlations. 
We also show that the finite-size scaling of the correlation ratio and spin structure factor is compatible with $\widetilde{\nu}=1/2$ and $\widetilde{\beta}=1/2$, as in the NP case.

However, our QMC results are not sufficient to establish the robustness of the nodal line against interactions. Although we observe small differences in the unequal-time Green's functions at different momenta along the nodal line, these differences cannot be reliably resolved in the spectral functions obtained through analytic continuation. Therefore, possible interaction-induced renormalizations of the low-energy band structure may remain below the momentum and energy resolution accessible in the present simulations.

We address this question by employing a renormalization-group analysis in the weak-coupling limit.
Our main result is that the nodal lines are unstable against interactions, ultimately leaving isolated nodal points at $k_z=\pm\pi/2$ along each original nodal line.
However, the associated renormalization of the band structure remains weak throughout the semimetallic regime. In particular, the interaction-induced gap at $\mathbf{k}=(\pi/2,\pi/2,0)$ remains smaller than $0.1t$, placing it at the limit of the energy resolution accessible in our QMC simulations.

Although small in energy, this reconstruction changes the dimensionality of the band-touching manifold, from nodal lines to isolated nodal points, with potentially important consequences for the critical behavior. In particular, the resulting low-energy spectrum may favor conventional three-dimensional mean-field criticality, providing a possible explanation for the exponents $\tilde{\nu}\approx1/2$ and $\tilde{\beta}\approx1/2$ obtained in our QMC simulations for the NL case. Nevertheless, given the strong finite-size effects discussed above, these numerical results should be treated as rough estimates.

In summary, our results suggest that, while the codimension of the nodal manifold controls the leading magnetic critical behavior (both the critical point and the critical exponents), the symmetries of the microscopic Hamiltonian determine whether the nodal manifold itself remains stable against interactions.
That is, we have highlighted the important role played by the geometry and codimension of the Fermi surface in interaction-driven quantum phase transitions.
We also emphasize that the instability of the nodal manifold is not a generic property of all symmetry-protected nodal-line semimetals, but rather of nodal-line realizations for which such a perturbation is permitted. If an additional crystalline symmetry forbids this bilinear, the nodal line may remain stable throughout the symmetry-preserving weak-coupling regime.

%%%%%%%%%%%%%%%%%%%%%%%%%%%%%%%%%%%%%%%%%%%%%%%%%%%%%%%%%%%%%%%%%%
\begin{acknowledgments}

The authors are grateful to H.~Guo, Z.-Y.~Meng, and R.~G.~Pereira for many insightful discussions on the subject.
A.R.M.S., L.O.L., M.B.S.N.~and N.C.C.~thank the Brazilian Agencies Conselho Nacional de Desenvolvimento Cient\'ifico e Tecnol\'ogico (CNPq), Coordena\c c\~ao de Aperfei\c coamento de Pessoal de N\'ivel Superior (CAPES), and Funda\c c\~ao de Amparo \`a Pesquisa do Estado do Rio de Janeiro, FAPERJ.
%\ncc{[Andressa, do you have your grant number?]}
A.R.M.S. thanks the International Institute of Physics of Federal University of Rio Grande do Norte (IIP-UFRN) and Funda\c c\~ao de Amparo \`a Pesquisa do Estado do Rio Grande do Norte (Grant No. 1699/24 IIF-FINEP).
M.B.S.N. acknowledges INCT -- Advanced Quantum Materials, involving CNPq (Proc.~408766/2024-7), FAPESP (Proc.~2025/27091-3), and CAPES.
N.C.C.~acknowledges support from FAPERJ Grant No.~E-26/210.592/2025, CNPq Grant No.~308130/2025-1, Serrapilheira Institute Grant No.~R-2502-52037, and the Alexander von Humboldt Foundation through a Humboldt Research Fellowship for Experienced Researchers.
F.F.A.~acknowledges  financial support by the Deutsche Forschungsgemeinschaft (DFG, German Research Foundation) through the W\"urzburg-Dresden Cluster of Excellence {\it ctd.qmat} -- Complexity, Topology and Dynamics in Quantum Matter (EXC 2147, Project No.~390858490).
The authors gratefully acknowledge the Gauss Centre for Supercomputing e.V. (www.gauss-centre.eu) for funding this project by providing computing time on the GCS Supercomputer SuperMUC-NG at Leibniz Supercomputing Centre
(www.lrz.de). 
\end{acknowledgments}

%%%%%%%%%%%%%%%%%%%%%%%%%%%%%%%%%%%%%%%%%%%%%%%%%%
\appendix

\section{Real-space Hamiltonian}
\label{app:real_space}
%%%%%%%%%%%%%%%%%%%%%%%%%%%%%%%%%%%%%%%%%%%%%%%%%%

A real-space representation is obtained by considering a cubic Bravais lattice with primitive vectors
$\mathbf{a}_1=a(2,0,0)$,
$\mathbf{a}_2=a(0,2,0)$,
$\mathbf{a}_3=a(0,0,1)$,
where $a$ is the lattice constant. The four orbitals inside the unit cell are located at
\begin{align}
\nonumber \mathbf{R}_A &= \mathbf{R},\\
\nonumber\mathbf{R}_B &= \mathbf{R}+\frac{\mathbf{a}_1}{2}+\frac{\mathbf{a}_2}{2},\\
\nonumber\mathbf{R}_C &= \mathbf{R}+\frac{\mathbf{a}_1}{2},\\
\mathbf{R}_D &= \mathbf{R}+\frac{\mathbf{a}_2}{2},
\end{align}
as shown in Fig.\,\ref{fig:lattice}.
By performing an inverse Fourier transform~\footnote{The Fourier transform of Eqs.~\eqref{eq:HamilrealspaceNP} and \eqref{eq:HamilrealspaceNL} gives Eq.\,\eqref{eq:H_toy}, with
$k_x\equiv \frac{\mathbf{k}\cdot\mathbf{a}_1}{2}$,
$k_y\equiv \frac{\mathbf{k}\cdot\mathbf{a}_2}{2}$,
$k_z\equiv \mathbf{k}\cdot\mathbf{a}_3$.
Thus \(k_x,k_y\in[-\pi/2,\pi/2)\), while \(k_z\in[-\pi,\pi)\).}, one finds the Hamiltonian for the nodal-point case as
\begin{align}
\label{eq:HamilrealspaceNP}
\nonumber \mathcal{H}_{0 ({\rm NP})}
& =
t\sum_{\mathbf{R}}
\biggl[
c^\dagger_{A,\mathbf{R}_{A}} c_{C,\mathbf{R}_{C}}
-
c^\dagger_{A,\mathbf{R}_{A}} c_{D,\mathbf{R}_{D}}
-
c^\dagger_{A,\mathbf{R}_{A}} c_{A,\mathbf{R}_{A}+\mathbf{a}_3}
\\
&
\nonumber +
c^\dagger_{C,\mathbf{R}_{C}} c_{A,\mathbf{R}_{A}+\mathbf{a}_1}
+
c^\dagger_{C,\mathbf{R}_{C}} c_{B,\mathbf{R}_{B}}
+
c^\dagger_{C,\mathbf{R}_{C}} c_{C,\mathbf{R}_{C}+\mathbf{a}_3}
\\
&
\nonumber +
c^\dagger_{D,\mathbf{R}_{D}} c_{B,\mathbf{R}_{B}}
+
c^\dagger_{D,\mathbf{R}_{D}} c_{D,\mathbf{R}_{D}+\mathbf{a}_3}
-
c^\dagger_{D,\mathbf{R}_{D}} c_{A,\mathbf{R}_{A}+\mathbf{a}_2}
\\
&
\nonumber +
c^\dagger_{B,\mathbf{R}_{B}} c_{D,\mathbf{R}_{D}+\mathbf{a}_1}
+
c^\dagger_{B,\mathbf{R}_{B}} c_{C,\mathbf{R}_{C}+\mathbf{a}_2}
-
c^\dagger_{B,\mathbf{R}_{B}} c_{B,\mathbf{R}_{B}+\mathbf{a}_3} \\
& + {\rm H.c.}
\biggr].
\end{align}
Here, $c^{\dagger}_{\alpha,\mathbf{R}_{\alpha}}$ and
$c_{\alpha,\mathbf{R}_{\alpha}}$ are, respectively, the creation
and annihilation operators for an electron in orbital $\alpha$ at
position $\mathbf{R}_{\alpha}$. The spin index has been omitted
for simplicity. The symbol ${\rm H.c.}$ denotes the
Hermitian conjugate of the preceding hopping terms.
Notice that  $\mathcal{H}_{0 ({\rm NP})}$ corresponds to the three-dimensional $\pi$-flux Hamiltonian, whose hopping pattern is shown in Fig.\,\ref{fig:lattice}.

The difference between the nodal-line and nodal-point limits is encoded in additional hopping processes. These terms are collected in $\mathcal{H}'$, so that the nodal-line Hamiltonian can be written as
$
\mathcal{H}_{0({\rm NL})}
=
\mathcal{H}_{0({\rm NP})}
+
\mathcal{H}'~,
$
with
\begin{align}
\label{eq:HamilrealspaceNL}
\nonumber \mathcal{H}'
& =
-\frac{t}{4}
\sum_{\mathbf{R}}
\biggl[
\sum_{\alpha=A,B}
\biggl(
c^\dagger_{\alpha,\mathbf{R}_{\alpha}}
c_{\alpha,\mathbf{R}_{\alpha}+\mathbf{a}_1+\mathbf{a}_3}
+
c^\dagger_{\alpha,\mathbf{R}_{\alpha}}
c_{\alpha,\mathbf{R}_{\alpha}+\mathbf{a}_1-\mathbf{a}_3}
\\
&
\nonumber +
c^\dagger_{\alpha,\mathbf{R}_{\alpha}}
c_{\alpha,\mathbf{R}_{\alpha}+\mathbf{a}_2+\mathbf{a}_3}
+
c^\dagger_{\alpha,\mathbf{R}_{\alpha}}
c_{\alpha,\mathbf{R}_{\alpha}+\mathbf{a}_2-\mathbf{a}_3}
\biggr)
\\
&
\nonumber -
\sum_{\alpha=C,D}
\biggl(
c^\dagger_{\alpha,\mathbf{R}_{\alpha}}
c_{\alpha,\mathbf{R}_{\alpha}+\mathbf{a}_1+\mathbf{a}_3}
+
c^\dagger_{\alpha,\mathbf{R}_{\alpha}}
c_{\alpha,\mathbf{R}_{\alpha}+\mathbf{a}_1-\mathbf{a}_3}
\\
&
 +
c^\dagger_{\alpha,\mathbf{R}_{\alpha}}
c_{\alpha,\mathbf{R}_{\alpha}+\mathbf{a}_2+\mathbf{a}_3}
+
c^\dagger_{\alpha,\mathbf{R}_{\alpha}}
c_{\alpha,\mathbf{R}_{\alpha}+\mathbf{a}_2-\mathbf{a}_3}
\biggr)
+
{\rm H.c.}
\biggr].
\end{align}

\section{Noninteracting density of states}
\label{app:dos}

The four-band Hamiltonian in Eq.\,\eqref{eq:H_toy_matrix} has the spectrum
\begin{equation}
E_{\pm}(\mathbf{k})
=
\pm 2t
\sqrt{
\cos^2 k_x+\cos^2 k_y+\chi^2(\mathbf{k})
},
\end{equation}
where each branch is twofold degenerate in orbital space. Including spin, each branch therefore has a total degeneracy of four.

For the nodal-point case, $\chi_{\rm NP}(\mathbf{k})=\cos k_z$, and the reduced Brillouin zone contains two inequivalent nodal points at $k_z=\pm\pi/2$. Expanding around either point gives an isotropic linear dispersion,
\begin{equation}
E_{\pm}(\mathbf{q})
\approx
\pm v |\mathbf{q}|,
\end{equation}
with $v=2t$.
The corresponding low-energy density of states is
\begin{equation}
\rho_{\rm NP}(E)
=
g_{\rm NP}
\int\frac{d^3q}{(2\pi)^3}
\delta\left(|E|-v|\mathbf{q}|\right)
=
g_{\rm NP}\frac{E^2}{2\pi^2v^3},
\end{equation}
with $g_{\rm NP}=2\times2\times2=8$, accounting for orbital degeneracy, spin, and the two nodal points. Hence,
$\rho_{\rm NP}(E) = \frac{E^2}{2\pi^2t^3}$, as presented in Eq.\,\eqref{eq:DOSNP}.

For the nodal-line case,
$\chi_{\rm NL}(\mathbf{k})
=
\cos k_z(\cos^2 k_x+\cos^2 k_y)$.
Expanding transverse to the nodal line gives
\begin{equation}
E_{\pm}(\mathbf{q}_\perp,k_z)
=
\pm 2t
\sqrt{
q_\perp^2+q_\perp^4\cos^2 k_z
}
\approx
\pm 2t q_\perp.
\end{equation}
Since the nodal line extends over the full $k_z$ direction, the low-energy density of states becomes
\begin{equation}
\rho_{\rm NL}(E)
=
g_{\rm NL}
\int_{-\pi}^{\pi}\frac{dk_z}{2\pi}
\int\frac{d^2q_\perp}{(2\pi)^2}
\delta\left(|E|-2tq_\perp\right),
\end{equation}
where $g_{\rm NL}=2\times2=4$ accounts for orbital degeneracy and spin. Performing the integrals gives $\rho_{\rm NL}(E) = \frac{|E|}{2\pi t^2}$, as presented in Eq.\,\eqref{eq:DOSNL}.

\section{Mean-field approximation}
\label{app:MFT}

We perform a static Hartree decoupling in the Hubbard term considering
\begin{align}
n_{\mathbf{R}_{\alpha}\uparrow}
n_{\mathbf{R}_{\alpha}\downarrow}
\approx{}&
\langle n_{\mathbf{R}_{\alpha}\uparrow}\rangle
n_{\mathbf{R}_{\alpha}\downarrow}
+
n_{\mathbf{R}_{\alpha}\uparrow}
\langle n_{\mathbf{R}_{\alpha}\downarrow}\rangle
\nonumber\\
&-
\langle n_{\mathbf{R}_{\alpha}\uparrow}\rangle
\langle n_{\mathbf{R}_{\alpha}\downarrow}\rangle,
\label{eq:MF_decoupling}
\end{align}
with $\alpha=A,B,C,D$ labeling the four sites within the unit cell.
For an arbitrary collinear magnetic state, we parametrize the local densities as
\begin{equation}
\langle n_{\mathbf{R}_{\alpha},\sigma}\rangle
=
\frac{n}{2}
+
\sigma \frac{m_{\mathbf{R}_{\alpha}}}{2},
\label{eq:MF_local_density}
\end{equation}
where $\sigma=+1$ for spin up and $\sigma=-1$ for spin down, while $n$ and $m$ define the local average density and the local average magnetization, respectively.
At half filling,
the spin-independent Hartree contribution can be absorbed into the chemical
potential.
Then, we assume a staggered antiferromagnetic pattern characterized by the wavevector
$\mathbf{Q}=(0,0,\pi)$, and the local magnetization
\begin{equation}
m_{\mathbf{R}_{A}}
=
m_{\mathbf{R}_{B}}
=
-m_{\mathbf{R}_{C}}
=
-m_{\mathbf{R}_{D}}
=
m \cos\left( {\mathbf{Q}\cdot\mathbf{R}} \right).
\label{eq:MF_AFM_ansatz}
\end{equation}

By performing a Fourier transform, as the magnetic order carries momentum $\mathbf{Q}$, the mean-field
term couples states at $\mathbf{k}$ and $\mathbf{k}+\mathbf{Q}$. Therefore, one should work with an enlarged spinor representation
\begin{equation}
\Phi_{\mathbf{k}\sigma}
=
\left(
\Psi_{\mathbf{k}\sigma},
\Psi_{\mathbf{k}+\mathbf{Q},\sigma}
\right)^T.
\end{equation}
The corresponding mean-field Hamiltonian can be written in block form as
\begin{equation}
\hat{H}_{\mathrm{MF},\sigma}(\mathbf{k})
=
\begin{pmatrix}
\hat{H}_{0}(\mathbf{k})
&
-\sigma \dfrac{Um}{2}\hat{\Lambda}
\\[6pt]
-\sigma \dfrac{Um}{2}\hat{\Lambda}
&
\hat{H}_{0}(\mathbf{k}+\mathbf{Q})
\end{pmatrix},
\label{eq:HMF_matrix}
\end{equation}
in which $\hat{H}_{0}(\mathbf{k})$ is the noninteracting Hamiltonian defined in Eq.\,\eqref{eq:H_toy_matrix}, and $\hat{\Lambda} = {\rm diag}(1,1,-1,-1)$.
The overall sign of the off-diagonal term can be changed by a redefinition
of the staggered order parameter and does not affect the spectrum.
Defining $\Delta=\frac{Um}{4t}$, and diagonalizing the resulting $8\times8$ matrix, one obtains the quasiparticle energies of Eq.\,\eqref{Eq:espectrum}.

The mean-field free energy is then obtained by summing over the
quasiparticle levels and adding the constant contribution generated by the
Hartree decoupling, which leads to Eq.\,\eqref{eq:MF_free_energy}.
The order parameter is obtained by imposing the stationary condition
$\partial F/\partial\Delta=0$. At finite temperature, the nontrivial
solution satisfies
\begin{equation}
\frac{t}{U}
=
\frac{1}{4N}
\sum_{\mathbf{k}}
\frac{
\tanh\left[
\beta t
\sqrt{
\cos^2 k_x+
\cos^2 k_y+
\chi^2(\mathbf{k})+
\Delta^2
}
\right]
}{
\sqrt{
\cos^2 k_x+
\cos^2 k_y+
\chi^2(\mathbf{k})+
\Delta^2
}
}.
\label{eq:MF_gap_finiteT_appendix}
\end{equation}
In the zero-temperature limit, the hyperbolic tangent approaches unity,
leading to
\begin{equation}
\frac{t}{U}
=
\frac{1}{4N}
\sum_{\mathbf{k}}
\frac{1}{
\sqrt{
\cos^2 k_x+
\cos^2 k_y+
\chi^2(\mathbf{k})+
\Delta^2
}
}.
\label{eq:MF_gap_T0_appendix}
\end{equation}

\section{Weak-coupling RG and Berry phase}
\label{app:RG}

We provide here the details of the weak-coupling RG analysis summarized
in Sec.~\ref{sec:RG}. The main point is that the onsite Hubbard
interaction is irrelevant at the Gaussian nodal-line fixed point, but
generates at second order a symmetry-allowed relevant quadratic
perturbation. Although parametrically small at weak coupling, this term
changes the asymptotic low-energy structure from a nodal line to
isolated nodal points.

Using
$\Gamma_1=\tau_x\otimes\mu_0$,
$\Gamma_2=\tau_y\otimes\mu_y$, and
$\Gamma_3=\tau_z\otimes\mu_0$,
the lattice nodal-line Hamiltonian can be written as
\begin{align}
\hat H_0(\mathbf{k})
=
2t\Big[
&\cos k_x\,\Gamma_1
+\cos k_y\,\Gamma_2
\nonumber\\
&+
\cos k_z
\left(
\cos^2k_x+\cos^2k_y
\right)\Gamma_3
\Big].
\label{eq:app_Hlattice}
\end{align}
Expanding around the nodal line,
$\mathbf{k}_{\rm NL}=(\pi/2,\pi/2,k_z)$, with
$k_x=\pi/2+q_x$ and $k_y=\pi/2+q_y$, gives
\begin{equation}
\hat H_0\approx -v(q_x\Gamma_1+q_y\Gamma_2)+\kappa q_\perp^2\cos k_z\,\Gamma_3,
\label{eq:app_Hcontinuum}
\end{equation}
where $v=\kappa=2t$ and $q_\perp^2=q_x^2+q_y^2$.
To leading order in $q_\perp$, the system may therefore be viewed as a
continuous family of $(2+1)$-dimensional Dirac cones parametrized by
$k_z$.

The Wilsonian RG transformation proceeds by integrating out modes in
the transverse momentum shell
$\Lambda/b<q_\perp<\Lambda$, with $b>1$, while the full $k_z$
dependence is retained. We parametrize the coarse graining by the
logarithmic running scale
\begin{equation}
\ell\equiv\ln b,
\qquad
\Lambda(\ell)=\Lambda_0e^{-\ell},
\label{eq:app_RG_scale}
\end{equation}
so that increasing $\ell$ corresponds to progressively lower
energies and momenta. Equivalently, along the RG flow,
$\omega(\ell)=\omega_0e^{-\ell}$ and
$\mathbf q_\perp(\ell)=\mathbf q_{\perp,0}e^{-\ell}$, whereas
$k_z$ is not rescaled. The corresponding running energy cutoff is
$E_\Lambda(\ell)=v\Lambda(\ell)=E_0e^{-\ell}$.

With this anisotropic scaling, the fermionic field has scaling
dimension $[\psi]=1$. Consequently, the onsite four-fermion coupling
has $[U]=-1$, whereas the coefficient $\Delta$ of the
symmetry-allowed operator $\Delta\cos k_z\,\Gamma_3$ has
$[\Delta]=+1$. The bare curvature coefficient is also irrelevant,
with $[\kappa]=-1$. Thus the Hubbard interaction is perturbatively
irrelevant at the Gaussian nodal-line fixed point, while a
momentum-independent $\Gamma_3$ term modulated by $\cos k_z$ is a
relevant perturbation.

\begin{figure}[t]
\centering
\includegraphics[width=0.9\linewidth]{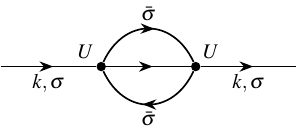}
\caption{
Second-order Hubbard self-energy responsible for the leading
momentum-dependent correction in the weak-coupling regime.
The first-order Hartree contribution vanishes at half filling in the
particle-hole-symmetric formulation. The $O(U^2)$ diagram generates,
upon projection onto the $\Gamma_3$ channel, the
$\cos k_z\,\Gamma_3$ perturbation discussed in the text.
}
\label{fig:RG_selfenergy}
\end{figure}

The Hubbard interaction is conveniently written in particle-hole
symmetric form as
\begin{equation}
H_U=U\sum_{\mathbf R,a}\left(n_{\mathbf R a\uparrow}-\frac12\right)
\left(n_{\mathbf R a\downarrow}-\frac12\right),
\label{eq:app_Hubbard}
\end{equation}
where $a=A,B,C,D$ labels the orbitals within the unit cell.
At half filling the first-order Hartree contribution vanishes.
The leading momentum-dependent correction is consequently the
second-order self-energy represented by
Fig.\,\ref{fig:RG_selfenergy}.

We denote fermionic and bosonic four-momenta by
$k=(\mathbf{k},i\omega_n)$ and
$q=(\mathbf{q},i\Omega_m)$, respectively, and use the notation
$\int_k\equiv(T/N)\sum_{\mathbf{k},\omega_n}$.
The bare propagator is the $4\times4$ orbital matrix
\begin{equation}
G^{(0)}(k)=\left[i\omega_n-\hat H_0(\mathbf{k})\right]^{-1}.
\label{eq:app_G0}
\end{equation}
For the local Hubbard vertex, the opposite-spin particle-hole bubble
connecting orbitals $a$ and $b$ is
\begin{equation}
\Pi^{(0)}_{ab,\bar\sigma}(q)=-\int_pG^{(0)}_{ab,\bar\sigma}(p+q)G^{(0)}_{ba,\bar\sigma}(p),
\label{eq:app_bubble}
\end{equation}
where the minus sign originates from the closed fermion loop.
The second-order self-energy may then be written as
\begin{equation}
\Sigma^{(2)}_{ab,\sigma}(k)=U^2\int_q\Pi^{(0)}_{ab,\bar\sigma}(q)G^{(0)}_{ab,\sigma}(k-q).
\label{eq:app_sigma_bubble}
\end{equation}
Equivalently, inserting Eq.\,\eqref{eq:app_bubble},
\begin{align}
\Sigma^{(2)}_{ab,\sigma}(k)
=
-U^2
\int_{p,q}
&
G^{(0)}_{ab,\sigma}(k-q)
G^{(0)}_{ab,\bar\sigma}(p+q)
\nonumber\\
&\times
G^{(0)}_{ba,\bar\sigma}(p).
\label{eq:app_sunset}
\end{align}
This is the orbital-space form of the second-order Hubbard
self-energy. The explicit orbital indices are important because the
interaction is local and diagonal in the orbital basis, whereas the
noninteracting propagator generally contains off-diagonal orbital
components.

%---------------------------------------------------------------------
% Second-order Hubbard self-energy
%---------------------------------------------------------------------

The component responsible for deforming the nodal manifold is obtained
by projecting the static self-energy onto $\Gamma_3$,
\begin{equation}
\Sigma_3^{(2)}(\mathbf{k},0)=\frac{1}{4}{\rm Tr}\left[\Gamma_3\Sigma^{(2)}(\mathbf{k},0)\right].
\label{eq:app_sigma3_projection}
\end{equation}
Along the original nodal line,
$\mathbf{k}=(\pi/2,\pi/2,k_z)$, its dependence on $k_z$ can be
expanded in lattice harmonics. The leading symmetry-allowed term is
the first harmonic,
\begin{equation}
\Sigma_3^{(2)}\left(\frac{\pi}{2},\frac{\pi}{2},k_z;0\right)=A_1\cos k_z+\cdots ,
\label{eq:app_sigma_harmonic}
\end{equation}
where the ellipsis represents higher harmonics and higher-order
interaction corrections. The coefficient $A_1$ is obtained by
projecting onto $\cos k_z$,
\begin{equation}
A_1=\frac{1}{\pi}\int_{-\pi}^{\pi}dk_z\,\cos k_z\,\Sigma_3^{(2)}\left(\frac{\pi}{2},\frac{\pi}{2},k_z;0\right).
\label{eq:app_A1}
\end{equation}
Evaluating the internal momentum and Matsubara sums with the full
lattice propagator gives
\begin{equation}
A_1=c_3\frac{U^2}{t},
\qquad
c_3\approx 2.91\times10^{-3}.
\label{eq:app_sigma_result}
\end{equation}
The leading interaction-generated correction to the single-particle
Hamiltonian is therefore
\begin{equation}
\delta\hat H_{\rm int}=\Delta_U\cos k_z\,\Gamma_3,
\qquad
\Delta_U=c_3\frac{U^2}{t}.
\label{eq:app_generated}
\end{equation}

We now incorporate this result into the Wilsonian flow. Introducing
the dimensionless Hubbard coupling $u$ and
$\delta=\Delta/E_\Lambda$, with $E_\Lambda=v\Lambda$, the
weak-coupling RG equations have the form
\begin{align}
\frac{du}{d\ell}
&=
-u+B_u u^2+\cdots,
\nonumber\\
\frac{d\delta}{d\ell}
&=
\delta+C_\Delta u^2+\cdots.
\label{eq:app_RGflows}
\end{align}
The first terms in these equations follow directly from tree-level
power counting. The $C_\Delta u^2$ contribution is the source term
associated with the second-order self-energy derived above. The
coefficients $B_u$ and $C_\Delta$ depend on the normalization of the
dimensionless couplings and on the cutoff convention, whereas the
coefficient $c_3$ obtained by matching to the full lattice
self-energy is a physical microscopic quantity.

In the weak-coupling regime, $u_0\ll1$, the nonlinear correction to
the first flow equation can be neglected, giving
\begin{equation}
u(\ell)=u_0e^{-\ell}.
\label{eq:app_u_solution}
\end{equation}
For the microscopic nodal-line Hamiltonian, $\delta_0=0$. Substitution
of Eq.\,\eqref{eq:app_u_solution} into the second flow equation yields
\begin{equation}
\delta(\ell)=\frac{C_\Delta u_0^2}{3}\left(e^\ell-e^{-2\ell}\right).
\label{eq:app_delta_solution}
\end{equation}
The interaction itself therefore flows to zero, whereas the relevant
quadratic perturbation generated by it grows under coarse graining.
This does not imply a divergent physical gap: since the running energy
cutoff behaves as $E_\Lambda=E_0e^{-\ell}$, the product
$E_\Lambda\delta(\ell)$ approaches a finite infrared value. Matching
to the lattice self-energy gives
$\Delta_{\rm phys}=\Delta_U=c_3U^2/t$.

The consequences for the nodal manifold follow immediately.
Along the original nodal line the effective Hamiltonian becomes
\begin{equation}
\hat H_{\rm eff}\left(\frac{\pi}{2},\frac{\pi}{2},k_z\right)=\Delta_U\cos k_z\,\Gamma_3,
\end{equation}
with spectrum
\begin{equation}
E_\pm(k_z)=\pm\left|\Delta_U\cos k_z\right|.
\label{eq:app_line_dispersion}
\end{equation}
Zero energy therefore survives only at
$k_z=\pm\pi/2$. Expanding around, for example,
$k_z=\pi/2+q_z$, for which $\cos k_z\approx -q_z$, gives
\begin{equation}
\hat H_{\rm NP}\approx -vq_x\Gamma_1-vq_y\Gamma_2-v_zq_z\Gamma_3,
\qquad
v_z\approx \Delta_U.
\label{eq:app_anisotropic_node}
\end{equation}
Thus the weak-coupling infrared theory is a strongly anisotropic
nodal-point semimetal, with
$v_z\propto U^2$, while $v_x\approx v_y\approx 2t$.
The crossover between nodal-line-like and nodal-point-like behavior
occurs at the parametrically small energy scale
$E_\times\sim\Delta_U\sim c_3U^2/t$.

%---------------------------------------------------------------------
\section{Berry phase and Wilson loop}
\label{app:berry}
%---------------------------------------------------------------------

The Berry-phase evolution provides a complementary geometrical characterization of this reconstruction. Since $\mu_y$ commutes with the Hamiltonian, the four-component orbital problem decomposes into two sectors $\eta=\pm1$, defined by $\mu_y|\eta\rangle=\eta|\eta\rangle$. In either sector the low-energy Hamiltonian becomes
\begin{equation}
h_\eta=-vq_x\tau_x-\eta vq_y\tau_y+M(k_z,q_\perp)\tau_z,
\label{eq:app_h_eta}
\end{equation}
with
\begin{equation}
M(k_z,q_\perp)=\cos k_z\left(\Delta_U+\kappa q_\perp^2\right).
\end{equation}

For $U=0$, the mass term vanishes as $q_\perp\rightarrow0$.
A transverse loop linking the nodal line therefore surrounds a two-dimensional Dirac point for every $k_z$, and each sector carries Berry phase $\gamma_\eta=\eta\pi$ modulo $2\pi$. Because the occupied orbital subspace is twofold degenerate, the gauge-invariant statement is that the corresponding Wilson-loop eigenvalues are both $-1$. This is the nodal-line counterpart of the $\pi$ Berry phase associated with an isolated Dirac point and of the band-contact-line picture discussed by Mikitik and Sharlai
\cite{MikitikSharlai1999,MikitikSharlai2008}.

For finite $U$, consider a circular transverse loop, $q_x=q_0\cos\phi$ and $q_y=q_0\sin\phi$. Neglecting the irrelevant curvature contribution $\kappa q_0^2$, the effective mass on a fixed-$k_z$ slice is $m(k_z)=\Delta_U\cos k_z$. The $\mathbf d$ vector of the corresponding two-level problem traces a circle of fixed polar angle on the Bloch sphere. Since the Berry phase is one half of the enclosed solid angle, one obtains
\begin{equation}
\gamma_\eta(k_z;q_0)=\eta\pi\left[1-\frac{\Delta_U\cos k_z}{\sqrt{v^2q_0^2+\Delta_U^2\cos^2k_z}}\right]
\quad
({\rm mod}\;2\pi).
\label{eq:app_Berry}
\end{equation}
For $U=0$, Eq.\,\eqref{eq:app_Berry} reduces to
$\gamma_\eta=\eta\pi$ independently of $k_z$. For finite interaction, the mass vanishes at the surviving nodes, $k_z=\pm\pi/2$, and the Berry phase remains exactly $\pi$ modulo $2\pi$ there. At a generic $k_z$, however, the corresponding two-dimensional slice is gapped, so the Berry phase of a finite loop is no longer quantized.

For displaying this evolution it is convenient to use the principal Wilson-loop phase,
\begin{equation}
\gamma_{\rm W}(k_z;q_0)=\pi\left[1-\frac{|\Delta_U\cos k_z|}{\sqrt{v^2q_0^2+\Delta_U^2\cos^2k_z}}\right],
\label{eq:app_Berry_principal}
\end{equation}
which takes values between $0$ and $\pi$. In the noninteracting case $\gamma_{\rm W}=\pi$ along the entire nodal line, whereas for $U\neq0$ the $\pi$ phase remains pinned only at $k_z=\pm\pi/2$. Moreover, in the limit $q_0\rightarrow0$ the Berry phase approaches zero modulo $2\pi$ for every gapped slice $k_z\neq\pm\pi/2$, while remaining $\pi$ at the two nodal points.

%Finally, the same interaction-generated term leads to a direct spectroscopic prediction. Along the original nodal line, the positive- and negative-energy branches are separated by
%
%\begin{equation}
%\Delta E(k_z)=2c_3\frac{U^2}{t}|\cos k_z|.
%\label{eq:app_spectral_prediction}
%\end{equation}
%
%The perturbative splitting is maximal near $k_z=0$ and $k_z=\pi$ and vanishes identically at $k_z=\pm\pi/2$. In particular, it is zero at $\mathbf{k}=(\pi/2,\pi/2,\pi/2)$, so the satellite structures observed at this momentum in Fig.\,\ref{fig:akw_pi2_NL}  \fa{Ref is not working!} cannot be attributed to the $O(U^2)\cos k_z\,\Gamma_3$ deformation obtained here. A direct test of the weak-coupling RG prediction would instead follow the low-energy spectral peaks along $(\pi/2,\pi/2,k_z)$ and determine whether their splitting exhibits the $U^2|\cos k_z|$ dependence of Eq.\,\eqref{eq:app_spectral_prediction}.

%%%%%%%%%%%%%%%%%%%%%%%%%%%%%%%%%%%%%%%%%%%%%%%%%%
% BIBLIOGRAPHY  
\bibliography{nl3d2}
%%%%%%%%%%%%%%%%%%%%%%%%%%%%%%%%%%%%%%%%%%%%%%%%%%
\end{document}